\documentclass{aa}  

\usepackage{graphicx}
\usepackage{txfonts}
\usepackage[colorlinks,allcolors=blue]{hyperref}

\usepackage{natbib}
\bibpunct{(}{)}{;}{a}{}{,} 
\usepackage{bm}
\usepackage{color}
\usepackage{ulem}
\usepackage{xspace}
\usepackage{float}
\usepackage{epstopdf}
\usepackage{CJKutf8}
\usepackage{orcidlink}

\ttfamily
\definecolor{mygray}{gray}{0.6}
\definecolor{magenta}{rgb}{0.858, 0.188, 0.478}
\definecolor{revise}{RGB}{0, 139, 0 }
\definecolor{revise2}{RGB}{255, 0, 0 }

\newcommand{\xxx}[1]{\textcolor{blue}{\textbf{xxx}\xspace}}

\newcommand{\fg}[1]{Fig.~\ref{fig:#1}}
\newcommand{\Fg}[1]{Figure~\ref{fig:#1}}

\newcommand{\eq}[1]{Eq.~(\ref{eq:#1})\xspace}
\newcommand{\eqb}[1]{Eq.~\ref{eq:#1}\xspace}

\newcommand{\tb}[1]{Table~\ref{tab:#1}\xspace}
\newcommand{\Tb}[1]{Table~\ref{tab:#1}\xspace}
\newcommand{\se}[1]{Sect.~\ref{sec:#1}\xspace}

\makeatletter
\def\uwave{\bgroup \markoverwith{\lower3.5\p@\hbox{\sixly \textcolor{red}{\char58}}}\ULon}
\font\sixly=lasy6 
\makeatother

\begin{document}

    \title{On the origin of transition disk cavities:}
    \subtitle{Pebble-accreting protoplanets vs Super-Jupiters}
    
    \author{Shuo Huang(\begin{CJK*}{UTF8}{gbsn}黄硕\end{CJK*}) \inst{1}\fnmsep\inst{2}
    \,\orcidlink{0000-0002-0054-8880}
    \and 
    Nienke van der Marel \inst{1} \,\orcidlink{0000-0003-2458-9756}
    \and 
    Simon Portegies Zwart \inst{1} \,\orcidlink{0000-0001-5839-0302}
    }
    \institute{Leiden Observatory, Leiden University, Einsteinweg 55, 2333 CC Leiden, The Netherlands
    \\
    \email{shuang@strw.leidenuniv.nl}
    \and
     Department of Astronomy, Tsinghua University, 100084 Beijing, China 
     }
    
    \date{Accepted XXX. Received YYY; in original form ZZZ}

 
  \abstract
   {Protoplanetary disks surrounding young stars are the birth places of planets. Among them, transition disks with inner dust cavities of tens of au are sometimes suggested to host massive companions. Yet, such companions are often not detected. 
   }
   {Some transition disks exhibit a large amount of gas inside the dust cavity and relatively high stellar accretion rates, which contradicts typical models of gas-giant-hosting systems. Therefore, we investigate whether a sequence of low-mass planets can create the appearance of cavities in the dust disk. 
   }
   {We evolve the disks with low-mass growing embryos in combination with 1D dust transport and {3D} pebble accretion, to investigate the reduction of the pebble flux at the embryos' orbits. We vary the planet and disk properties to understand the resulting dust profile.
   }
   {We find that multiple pebble-accreting planets can efficiently decrease the dust surface density, resulting in dust cavities consistent with transition disks. The number of low-mass planets necessary to sweep up all pebbles decreases with decreasing turbulent strength and is preferred when the dust Stokes number is $10^{-2}-10^{-4}$. Compared to dust rings caused by pressure bumps, those by efficient pebble accretion exhibit {more extended outer} edges. We also highlight the observational reflections: the transition disks with rings featuring {extended outer} edges tend to have a large gas content in the dust cavities and rather high stellar accretion rates. 
   }
   {We propose that planet-hosting transition disks consist of two groups. In Group A disks, planets have evolved into gas giants, opening deep gaps in the gas disk. Pebbles concentrate in pressure maxima, forming dust rings. In Group B, multiple Neptunes (unable to open deep gas gaps) accrete incoming pebbles, causing the appearance of inner dust cavities and distinct ring-like structures near planet orbits. The morphological discrepancy of these rings may aid in distinguishing between the two groups using high-resolution ALMA observations. 
   }

   \keywords{Protoplanetary disk -- Methods: numerical -- planet formation}

   \maketitle
%

\section{Introduction}
Since the discovery of the first exoplanet around a solar-type star \citep{MayorQueloz1995}, thousands of mature exoplanets have been confirmed, which keep challenging our understanding of planet formation. Planets are expected to form in protoplanetary disks. Thanks to the Atacama Large Millimeter/submillimeter Array (ALMA), hundreds of planet-forming disks have been imaged and studied \citep[e.g., ][]{AndrewsEtal2018, LongEtal2018}. 


A group of disks of particular interest are classified as "transition disks", featuring large cavities in the dust continuum \citep[e.g.][and references therein]{KenyonHartmann1995,vanderMarel2023}. Such large cavities in dust disks can indicate the existence of planets. These cavities are potentially cleared by super Jovian planets, as they are able to perturb the gas disk and create pressure bumps \citep{LinPapaloizou1979, GoldreichTremaine1980}, leading to the convergence of dust and the formation of distinct ring-like structures within the disks. 

Some transition disks, including PDS 70, have much lower stellar accretion rates than typical full disks with similar stellar mass as well as a deep inner gas cavity, as demonstrated by spatially resolved $^{12}$CO and $^{13}$CO images \citep{MuleyEtal2019, FacchiniEtal2021}. Disks with giant planets (such as PDS 70) may have already evolve to the rather later stage of giant planet formation. Other transition disks have more typical accretion rates, indicating material transport across the gap. Possibly, 
their planet's masses are not so high that only shallow gas gaps are opened \citep{Dodson-RobinsonSalyk2011, LeemkerEtal2022}. 

It remains a question whether there are super-Jovian planets in most of the transition disks. 
The confirmation of proto-planets through direct imaging within the circumstellar disks was achieved only in three disks: PDS 70 \citep{KepplerEtal2018, HaffertEtal2019}, AB Aur \citep{CurrieEtal2022, ZhouEtal2022} and HD 169142 \citep{HammondEtal2023}, whereas most other targets do not show significant planet signals \citep{Asensio-TorresEtal2021}. 



Considering the lack of direct confirmation of massive companions in the disk gaps and cavities as well as the difficulty of explaining giant planet formation on wide orbits, various other mechanisms have been proposed to explain gaps and cavities. 
With pressure bumps, these mechanisms include disk dispersal through MHD-driven winds \citep{TakahashiMuto2018} or photoevaporative flows \citep{ErcolanoPascucci2017}, zonal flows in MHD disks \citep{JohansenEtal2009}, mass pile-up at the boundary between magnetically active and dead zones \citep{FlockEtal2015}, and spontaneous ring formation due to reduced accretion by concentrated dust \citep{DullemondPenzlin2018, HuEtal2019}.
Moreover, rings can form even without the presence of pressure maxima due to the condensation fronts at icelines \citep{ZhangEtal2015}, temperature dips \citep{ZhangEtal2021i}, traffic jam effect \citep{JiangOrmel2021}, and grain growth processes \citep{OhashiEtal2021}.

Recent studies have revisited the possibility that most, if not all cavities are caused by planets. On one hand, pebble accretion has been shown to efficiently grow planets even in orbits wider than 10 au \citep{JohansenBitsch2019, JiangOrmel2023, GurrutxagaEtal2024, HuangEtal2024i}. Also, the occurrence rate of giant planets exhibits a similar dependence on stellar mass as the occurrence rate of structured protoplanetary disks \citep{vanderMarelMulders2021}. This suggests that the disk substructures could be primarily caused by planets.

Low-mass planets are not often considered as the initiator of disk substructures, since they cannot generate pressure bumps.
However, they may still alter the dust transport. In disks hosting low-mass planets, dust tends to drift inward. When the inward pebble flux intersects with a planet's orbit, some of it can be accreted onto the planet's surface — a process known as pebble accretion \citep{OrmelKlahr2010, LambrechtsJohansen2012}. Consequently, the planet acts as a filter, reducing the inward pebble flux. In this paper, we investigate whether multiple planets, from Neptunian up to Saturnian mass, can lead to the formation of dust gaps and cavities due to pebble accretion only, without the aid of gas pressure bumps.

The paper is structured as follows: We present the dust transport model and simulation initial conditions in \se{method}. In \se{main}, we investigate and analyze dust transport and its dependence on planet and disk properties. Following this, in \se{alma}, we compare the morphology differences between the rings induced by Neptunian- and (super-)Jupiter-mass planets and find observational consistencies with our theoretical understanding. We discuss and summarize the results in \se{discussion} and \se{conclusion}, respectively. 

\section{Methods}
\label{sec:method}
\subsection{Dust transport including pebble accretion}
The dust surface density during its transport is modeled through the 1D advection-diffusion equation \citep[e.g.][]{BirnstielEtal2012}:
\begin{equation}
\label{eq:diffusion}
    \frac{\partial \Sigma_\mathrm{d}}{\partial t}+
    \frac{1}{r}\frac{\partial}{\partial r}(rv_\mathrm{drift}\Sigma_\mathrm{d})-
    \frac{1}{r}\frac{\partial}{\partial r}\left[rD\frac{\Sigma_\mathrm{g}}{\Sigma_\mathrm{d}}\frac{\partial}{\partial r}\left(\frac{\Sigma_\mathrm{d}}{\Sigma_\mathrm{g}}\right)\Sigma_\mathrm{d}\right]=0,
\end{equation}
where $\Sigma_\mathrm{d}$ and $\Sigma_\mathrm{g}$ are the dust and gas surface densities, respectively, and $D$ is the dust diffusivity. The dust loses angular momentum and drifts inward, due to the friction between the gas in the sub-Keplerian rotation and the dust in the Keplerian rotation. The dust radially drifts \citep{NakagawaEtal1986} at the speed of:
\begin{equation}
    v_\mathrm{drift}=-\frac{2\mathrm{St}}{1+\mathrm{St}^2}\eta v_\mathrm{K},
\end{equation}
where $\mathrm{St}$ and $v_\mathrm{K}$ represent dust Stokes number and local Keplerian velocity, respectively. The mid-plane gas pressure gradient $\eta$ is expressed as:
\begin{equation}
\label{eq:eta}
    \eta=-\frac{1}{2}\frac{\partial\log{P}}{\partial\log{r}}\left(\frac{H_\mathrm{g}}{r}\right)^2,
\end{equation}
and $H_\mathrm{g}=c_s r/v_\mathrm{K}$ is the gas scale height, $c_s=\sqrt{k_\mathrm{B}T/\mu m_H}$ is the sound speed at the mid-plane. The gas mean molecular weight is taken to be 2.3.

We subtract the accreted pebble flux to account for pebble accretion during dust transport. The numerical treatment is given in \se{numerical}. We calculated the 3D pebble {accretion} efficiency following \cite{OrmelLiu2018}. {This calculation depends on the planet's eccentricity, the dust Stokes number $\mathrm{St}$, the local gas pressure gradient $\eta$, the gas scale height $H_\mathrm{g}$, and the vertical viscosity. The radial diffusion of the dust particles is ignored.}

The diffusion process is ignored when including pebble accretion (\se{main} and \se{gmaur}). The steep pebble profile driven by pebble-accreting planets induces unrealistic leaks of pebbles near planet orbits in the numerical simulation ($D\frac{\Sigma_\mathrm{g}}{\Sigma_\mathrm{d}}\frac{\partial}{\partial r}\left(\frac{\Sigma_\mathrm{d}}{\Sigma_\mathrm{g}}\right)\gg -v_\mathrm{drift}$ in \eqb{diffusion}). The consequence of such simplification is discussed in \se{assessment}. {We also assess that dust radial diffusion can be negligible in the process of pebble accretion in \se{diffusion}, as the distance dust particles diffuse is smaller than the planet's Hill radius over one synodical timescale.}


\subsection{Initial conditions}
Since the dust radial drift timescale is much shorter than the disk typical disk lifetime, the gas profile is fixed over time. We set up a protoplanetary disk with the usual power law plus cutoff shape \citep{Lynden-BellPringle1974}, 
\begin{equation}
    \Sigma_\mathrm{g}(r) = \frac{M_\mathrm{d}}{2\pi R_\mathrm{out}(1-e^{-1})}\frac{\exp(-r/R_\mathrm{out})}{r},
\end{equation}
where $M_\mathrm{d}=0.1M_\star$ is the disk mass and $R_\mathrm{out}$ is the disk characteristic radius. 
The initial dust surface density distribution is set after assuming an initially constant dust-to-gas ratio:
\begin{equation}
    \Sigma_\mathrm{d}(t=0,r)=Z\Sigma_\mathrm{g}(r)
\end{equation}
where $Z$ stands for the metallicity. Then the dust can be transported in the fixed background of gas. 

The gas temperature profile is
\begin{equation}
    T(r) = 200K\left(\frac{r}{1\,\mathrm{au}}\right)^{-0.5},
\end{equation}
which also sets the gas scale height, as well as the dust scale height through: 
\begin{equation}
\label{eq:scaleheight}
    H_\mathrm{d}=\left(1+\frac{\mathrm{St}}{\alpha}\frac{1+2\mathrm{St}}{1+\mathrm{St}}\right)^{-1/2}H_\mathrm{g}
\end{equation}
\citep{YoudinLithwick2007}, where $\alpha$ is the gas viscosity parameter and St is the dust Stokes number. The St is defined by $\mathrm{St}=\pi s \rho_\mathrm{s}/2\Sigma_\mathrm{g}$, where $s$ and $\rho_\mathrm{s}$ are the dust grain size and internal density.
The choices of different temperature profiles hardly affect our calculations, which is discussed further in \se{assessment}. 

\section{Pebble flux reduction}
\label{sec:main}
In this section, we study how much the Neptunian- to Saturnian-mass planets (which are too small to open gas gaps $\gtrsim10$ au at our disk conditions) reduce the pebble flux via pebble accretion around a solar-type star without making pressure bumps. The simulations are conducted using AMUSE \citep{PortegiesZwartEtal2009, PortegiesZwartEtal2013, PelupessyEtal2013}. 

\subsection{Pebble accretion efficiency}
\label{sec:peb_acc}
Once a planet grows sufficiently large to induce a pressure bump exterior to its orbit, it halts the incoming pebble flux and, as a consequence, pebble accretion stops. The maximum mass that one planet can reach before opening a gap is the pebble isolation mass, which is fitted from 3D-hydrodynamical calculation \citep{BitschEtal2018}:
\begin{equation}
\begin{aligned}
    m_\mathrm{iso} = & 25 M_\oplus \left(\frac{h}{0.05}\right)^3\left[0.34\left(\frac{-3}{\log_{10}\alpha}\right)^4+0.66\right] \\
    & \times \left[1- \frac{\partial \ln P/\partial \ln r+2.5}{6}\right]\frac{M_\star}{M_\odot}.
\end{aligned}
\end{equation}
Here $\partial P/\partial r$ is the local pressure gradient of the radial gas disk. Planets at pebble isolation mass can decrease the ambient gas density by 10-20\% \citep{BitschEtal2018}. The pebble isolation mass at different distances with different viscosities is shown in \fg{isolation}. The pebble isolation mass increases towards high $\alpha$ and longer distance. 

\begin{figure}[ht!]
    \centering
    \resizebox{\hsize}{!}{\includegraphics[width=\columnwidth]{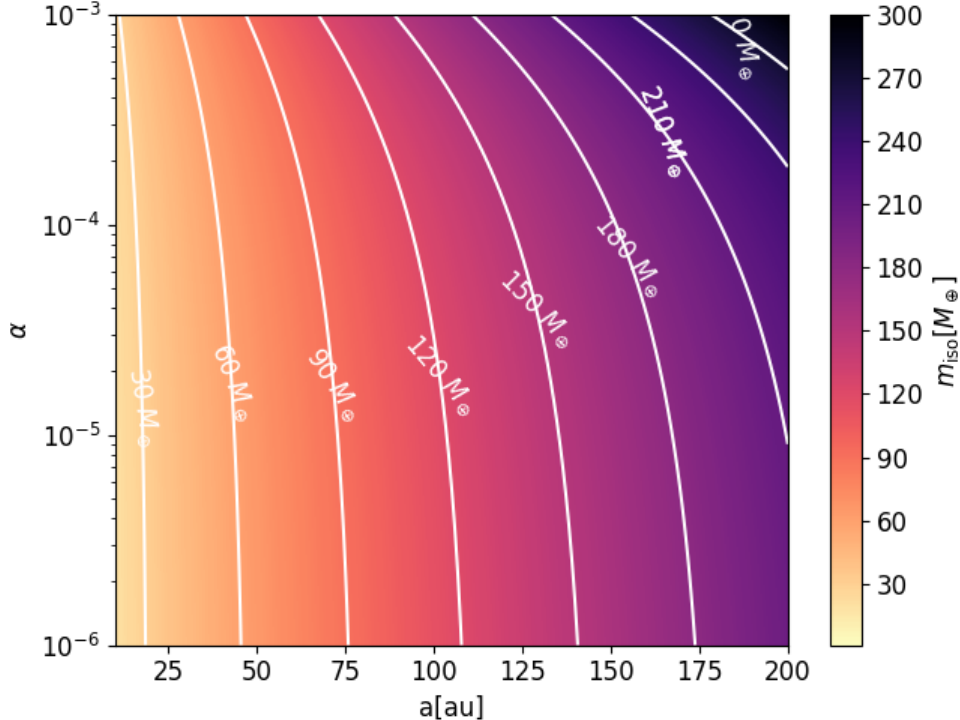}}
    \caption{Pebble isolation mass around a solar-type star, with varying viscosity-$\alpha$ (y-axis) at different orbits (x-axis). The color bar indicates the pebble isolation mass calculated following \cite{BitschEtal2018}. }
    \label{fig:isolation}
\end{figure}

We calculate the pebble accretion efficiency $\epsilon_\mathrm{PA}$, assuming planets are in circular orbits. Remarkably, eccentric orbits (10$^{-2}$-10$^{-1}$) can enhance the pebble accretion efficiency by a factor of up to {3-5} \citep{LiuOrmel2018}. The planet mass is fixed at the local pebble isolation mass. 

Our calculation in \fg{efficiency} shows {the same results as \cite{OrmelLiu2018} found}. When $\alpha$ increases, the pebble accretion efficiency decreases. That is because higher $\alpha$ stirs up more pebbles, especially with low $\mathrm{St}$. This lowers $\epsilon_\mathrm{PA}$ as 3D pebble accretion becomes more common. Efficiency increases and then decreases as $\mathrm{St}$ grow from $10^{-5}$ to $10^{-1}$. The accretion efficiency peaks when the pebble accretion cross section matches the pebble height, signaling the switch from 2D to 3D pebble accretion. As $\alpha$ drops, the peak shifts down because the dust height decreases. Bigger pebbles (higher $\mathrm{St}$) lose efficiency since they drift too fast, missing the planet. Smaller pebbles (lower $\mathrm{St}$) have a better chance of being captured because they spread out more.

\begin{table*}
    \centering
    \begin{tabular}{c|c|l}
        \hline
        {Parameters}  &  {Value(s)} & {Description}\\
        \hline
        $\alpha$ & $10^{-6}$, $10^{-5}$, $10^{-4}$ & Viscosity parameter \\
        $\mathrm{St}$ & $10^{-3.5}$, $10^{-3}$, $10^{-2.5}$ & Stokes number of dust \\
        $m_\mathrm{p}/m_\mathrm{iso}$ & 0 (non-planet case), 0.5, 0.75, 1.0 & Planet mass normalized by local pebble isolation mass \\
        \hline
        $Z$ &  0.0134 & Disk metallicity \\
        $M_\mathrm{d}$ & 0.1$M_\star$ & Disk mass \\
        $R_\mathrm{out}$ & 200 au & Disk characteristic radius\\
        $\mu$  & 2.3 & Gas mean molecular weight  \\
        $M_\star$ & $M_\odot$ & Star mass \\
        \hline
    \end{tabular}
    \caption{ Definition of quantities, symbols, and units. The first three parameters separated from the rest by a horizontal line are the free parameters of the simulations. }
    \label{tab:parameter}
\end{table*}

\begin{figure*}
    \centering
    \includegraphics[width=1.8\columnwidth]{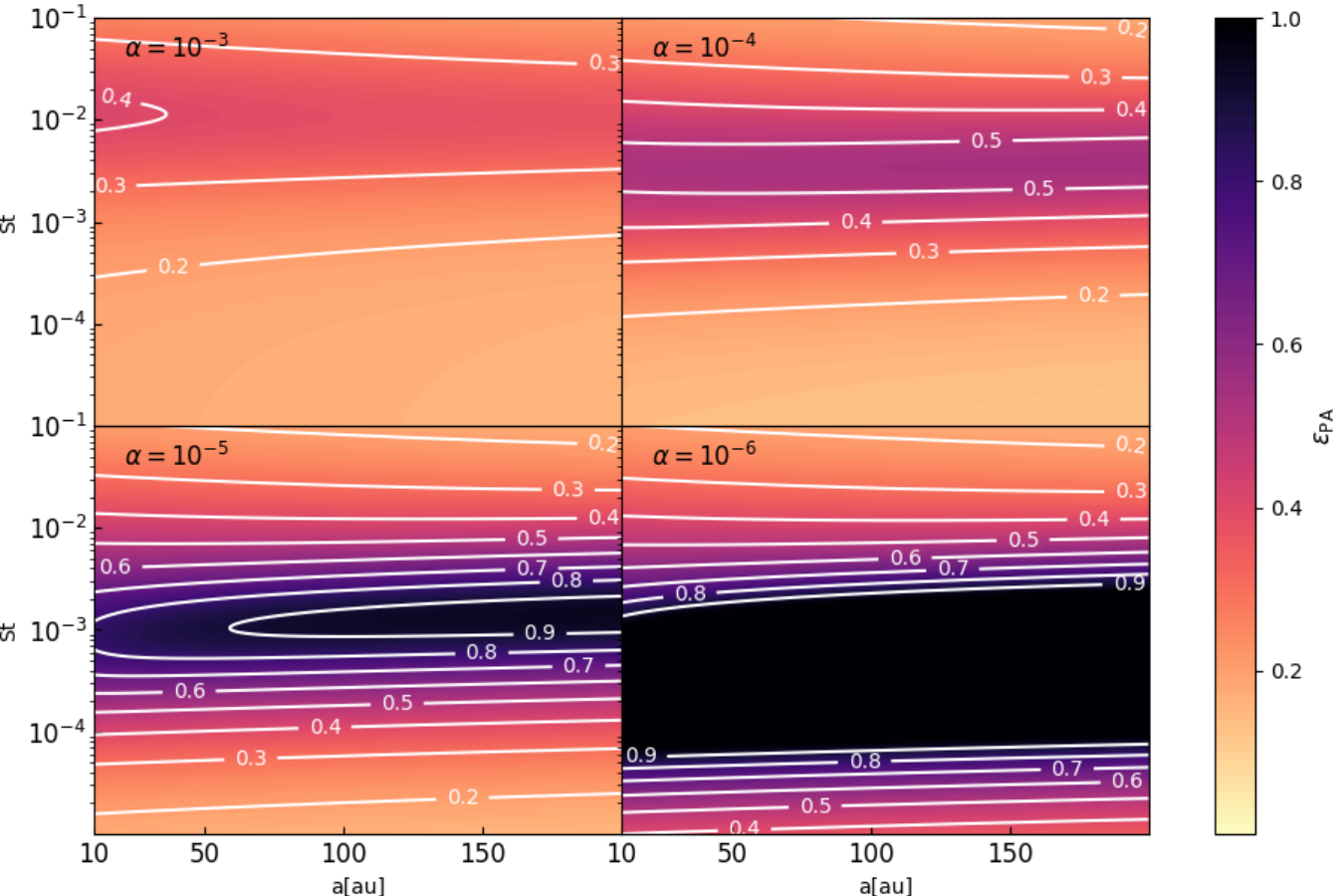}
    \caption{Efficiency map illustrating pebble accretion onto a planet in a circular orbit. We manipulate the planet's semimajor axis (x-axis) and the dust Stokes number (y-axis). The planet's mass corresponds to the pebble isolation mass within its local disk. Variations in viscosity ($\alpha$) are indicated across four panels. Contours and the color display the values of the pebble accretion efficiency. }
    \label{fig:efficiency}
\end{figure*}

\begin{figure*}
    \centering
    \includegraphics[width=2\columnwidth]{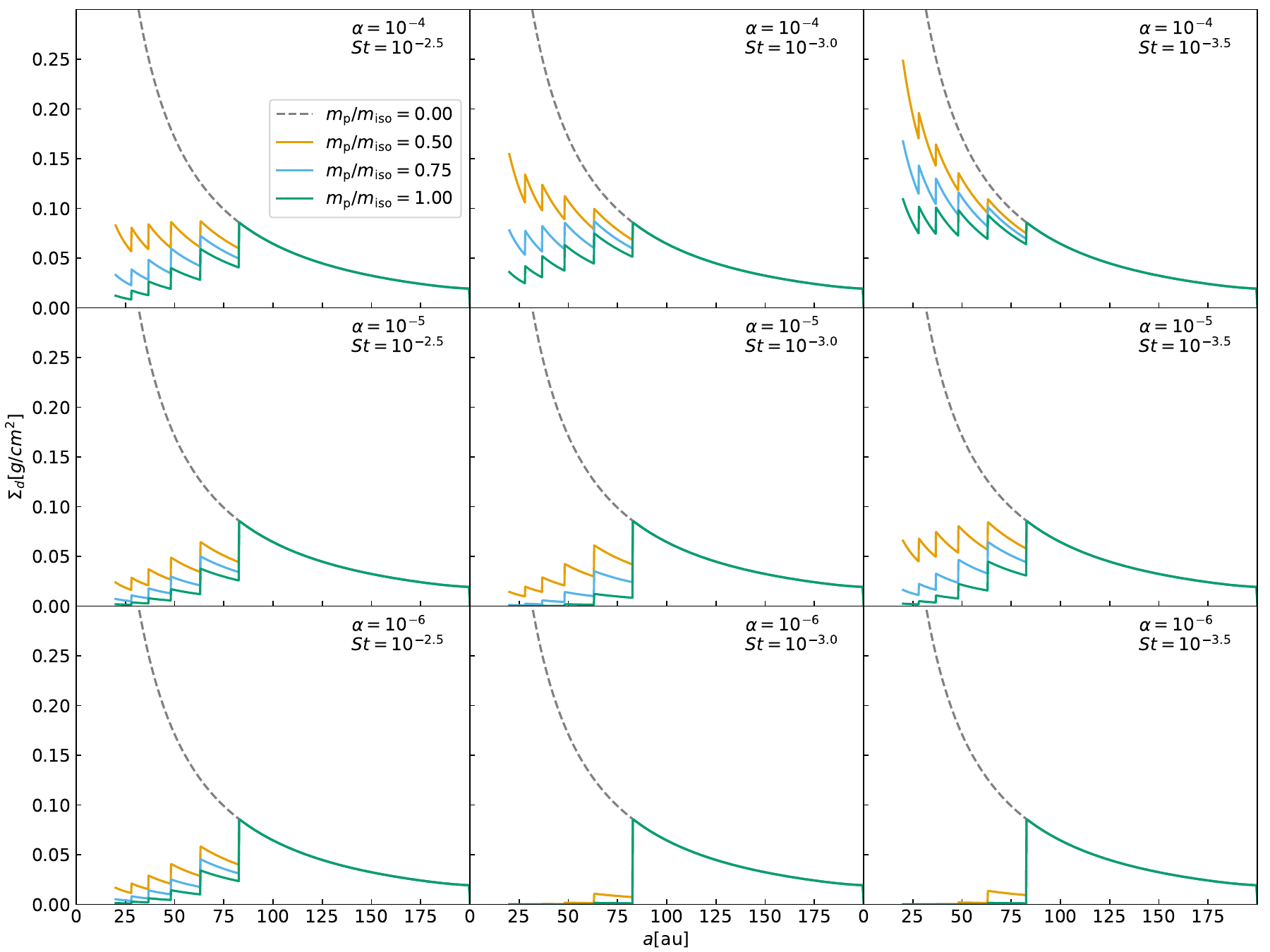}
    \caption{Equilibrium dust surface density profiles across various planet masses, viscosities, and Stokes numbers at different planetary orbits. There are five planets in a 3:2 MMR chain and dust surface density drops at planets' orbits due to pebble accretion. Each row corresponds to a different $\alpha$ value, while each column represents a different Stokes number ($\mathrm{St}$). The color distinguishes between dust profiles associated with different planet masses, with the grey dashed line denoting the no-planet case.}
    \label{fig:dust_profile}
\end{figure*}

As $\alpha$ drops below $10^{-4}$, the peak $\epsilon_\mathrm{PA}$ reaches 0.5. This sweet spot expands as $\alpha$ decreases. At $\alpha=10^{-5}$, $\epsilon_\mathrm{PA}$ exceeds 0.9 around $\mathrm{St}=10^{-3}$ beyond 50 au. With $\alpha=10^{-6}$, $\epsilon_\mathrm{PA}$ hits unity if $\mathrm{St}$ falls between $6\times10^{-5}$ and $2\times10^{-3}$ at nearly all distances in the disk. The accretion sweet spot (with high $\epsilon_\mathrm{PA}$) area expands even more if lower $\alpha$ is used.

Pebble accretion is very efficient, with an efficiency always over 0.1, even at $\alpha=10^{-3}$. This is because the planet mass we used is the maximum mass that the planet can reach before opening a deep gas gap.
The pebble accretion efficiency for smaller planets can be roughly estimated. It shows a positive correlation with planet mass, following a power law ranging from 0.5 to 1.0 \citep[in the 2D and 3D regime, respectively][]{OrmelLiu2018}. For example, a planet with half the pebble isolation mass exhibits roughly half the accretion efficiency depicted in \fg{efficiency}.

    

\subsection{Dust cavity and ring structure}
Efficient pebble accretion can shape the dust continuum profile noticeably due to planets. This effect is expected to be more apparent in closely packed multi-planet systems. In this section, we focus on understanding the geometry of the resulting dust substructures and how they relate to disk and planet parameters.

Due to the angular momentum exchange between the gas disk and planets, planets can migrate in the disk \citep{LinPapaloizou1979, GoldreichTremaine1980}. Mean Motion Resonance (MMR) is a natural outcome of planet migration \citep{TerquemPapaloizou2007, RaymondEtal2008}. This is consistent with exoplanet observations as we already find many multi-planet systems in resonance chains. For instance, the two \citep[possibly three][]{ChristiaensEtal2024i} planets in PDS 70 are in MMR(s) \citep{WangEtal2021i}. The four directly imaged planets in HR 8799 are in MMRs \citep{ZurloEtal2022}. There are more planets in MMRs in the Kepler field \citep[10-20\%, ][]{HuangOrmel2023i, HamerSchlaufman2024}.  

Therefore, we begin with {five} planets arranged in a 3:2 resonance pattern, with their orbital periods close to integer ratios. The innermost planet is positioned at 28 au, with subsequent planets at 36.7, 48.1, 63.0, and 82.6 au. The mutual dynamical interactions are not modeled, instead, we assume they are stable at their orbits and then study the disk evolution using the 1D model. Simulations run until $t_\mathrm{end}=0.5 (0.01/\mathrm{St})$ Myr, allowing sufficient drift of dust particles with varying $\mathrm{St}$. Extending the simulation duration does not alter the normalized density distribution's shape, only decreasing its absolute values.

We vary three parameters: planet mass, disk viscosity ($\alpha$), and the dust Stokes number $\mathrm{St}$. All the planets have the same mass ratio $m_\mathrm{p}/m_\mathrm{iso}$, reflecting the ratio of the planet's mass to the local pebble isolation mass. The parameter values/ranges are shown in \tb{parameter}. The ranges of $\alpha$ and St are further assessed in \se{sweetspot}. 

The resulting dust profile is shown in \fg{dust_profile}. Reductions in dust surface density appear at all planetary orbits due to pebble accretion. Overall, the inward pebble flux decreases at each of the planet locations, resulting in lower surface densities compared to scenarios without planets. Notably, most disks in the simulations with planets show inner cavity features, with their dust surface density peaking at $\approx80$ au, far from the disk's inner edge.

Our simulations are consistent with the efficiency map of pebble accretion discussed in \se{peb_acc}. The depth of the cavity increases as $\alpha$ decreases, corresponding to the rise in $\epsilon_\mathrm{PA}$. Furthermore, when $\alpha$ is fixed, the deepest cavity shifts towards smaller $\mathrm{St}$ values, mirroring the movement of the high-efficiency sweet spot indicated by the deep-colored regions in \fg{efficiency}.

When the mass ratio $m_\mathrm{p}/m_\mathrm{iso}=1.0$, the simulation with $\alpha=10^{-4}$ and {$\mathrm{St}=10^{-3.5}$} is particularly interesting. The simulation does not exhibit the cavity feature. Instead, the peak dust surface density is near the disk's inner edge. This occurs because of the combination of small dust particles and high viscosity, resulting in a dust scale height much greater than the planet's accretion cross-section. Consequently, the efficiency of pure 3D pebble accretion is insufficient to create a cavity-like feature in the dust profile with five planets.

The fraction of disks with a cavity is low in the simulations when the planets are relatively small $m_\mathrm{p}/m_\mathrm{peb}=0.5$. Essentially, the cross-section between the pebbles and planets is proportional to the planet Hill radius \citep{OrmelLiu2018}, hence smaller planets induce a lower net probability of pebble accretion and shallow cavity. In theory, a larger number of planets in the system could still result in the formation of a cavity, with the depth of the cavity increasing as the number of planets rises. 

If the viscosity is at its minimum in our parameter space ($\alpha=10^{-6}$), the deep cavity persists. The cavity radius reaches $\approx80$ au, at the orbit of the outermost planet when $\mathrm{St}=10^{-3}$ and $10^{-3.5}$, because the efficiency of planet pebble accretion approaches unity with low turbulence. In such cases, the outermost one or two planets can accrete the entire incoming pebble flux, leading to the formation of the inner cavity. Lower viscosity facilitates the creation of deep cavities even in systems with fewer planets.

As the disk evolves and approaches the stage where the dust disk becomes optically thin, the continuum intensity becomes proportional to the dust surface density. Consequently, density drops at the orbits of planets manifesting as ring-like structures. These rings exhibit a distinctive feature: the inner part of the ring appears sharp due to efficient pebble accretion, while the outer part of the ring appears much smoother, as outer pebbles gradually and smoothly drift inward. This special feature can be used to distinguish pebble accretion-induced rings from others in observations, especially when the disk substructures can be resolved to spatial scales of $\lesssim10$ au with ALMA at high angular resolution.

\begin{table*}
    \centering
    \caption{ Source properties and information of the used ALMA observation. }
    \begin{tabular}{c|c|c|c|c|c|c|l}
        \hline
        {Source}  &  {Inclination} & {Position Angle\tablefootmark{a}[$^\mathrm{o}$]} & {Distance[pc]} & {FWHM[mas]} & PA\tablefootmark{b}[$^\mathrm{o}$] & Band & Reference \\
        \hline
        GM Aur  & 53.21 & 57.17 & 159 & 45$\times$25 & 2.2 & 6 & \cite{HuangEtal2020} \\
        PDS 70  & 51.7  & 156.7 & 112.4 & 74$\times$57 & 63 & 7 & \cite{KepplerEtal2018} \\

        \hline        
    \end{tabular}
    \tablefoot{\tablefoottext{a}{The angle between the major axis of the disk to the North.}\\
        \tablefoottext{b}{The angle between the major axis of the beam to the North.}}
        
    \label{tab:source}
\end{table*}
\begin{figure*}
    \sidecaption
    \includegraphics[width=1.1\columnwidth]{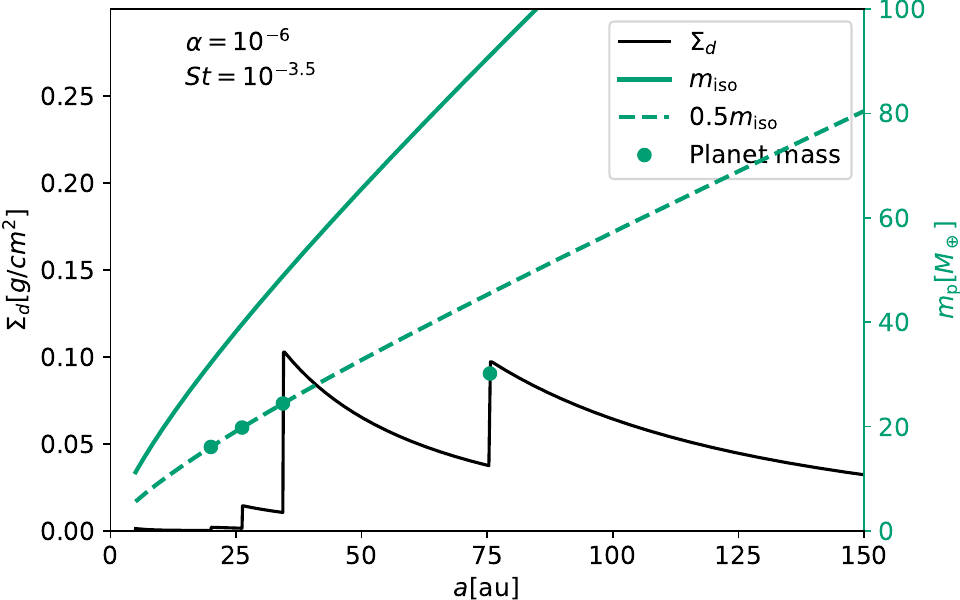}
    \caption{Dust profile and the planet mass in the simulated GM Aur disk. The disk parameters used are labeled on the top left. The solid line is the dust surface density profile and the values are shown on the left y-axis. The black dashed line represents the pebble isolation mass at different locations (x-axis), while the grey dashed line represents half of the pebble isolation mass. Black dots indicate the location and the mass of the planets. The right y-axis shows the values of planet mass. }
    \label{fig:GMaurprofile}
\end{figure*}

\begin{figure*}
    \sidecaption
    {\includegraphics[trim={0 0 2.7cm 0},clip, height=6cm]{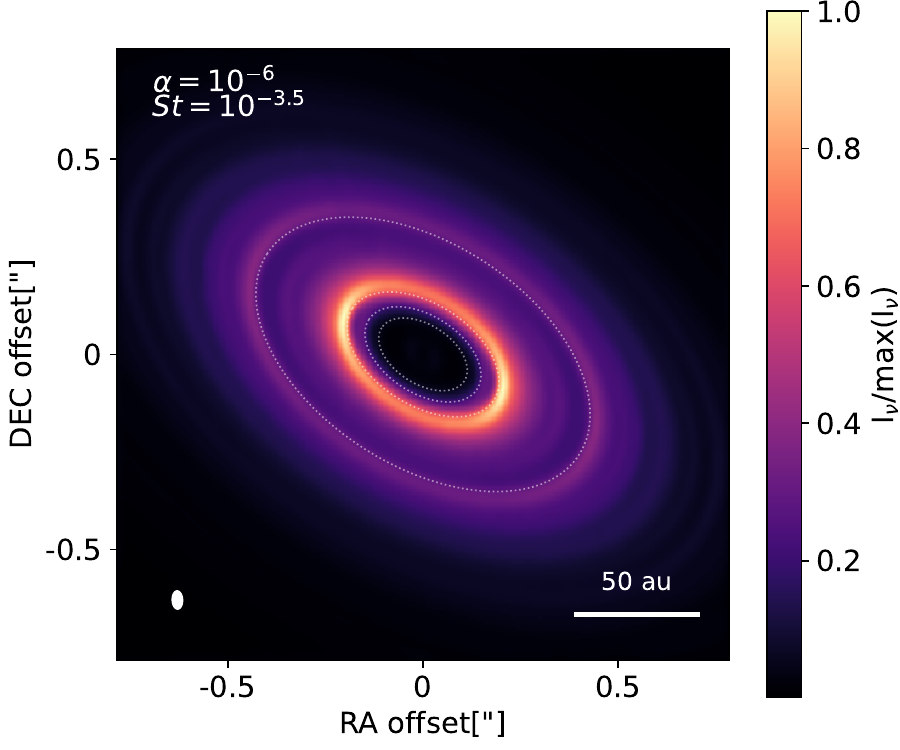}
    \includegraphics[trim={1.8cm 0 0 0},clip, height=6cm]{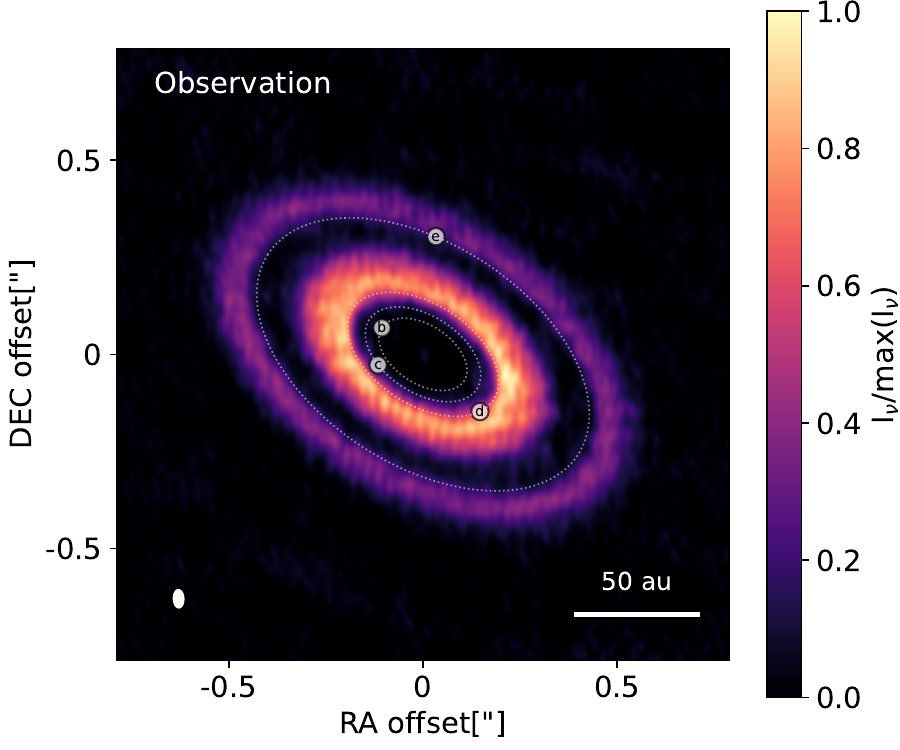}}
    \caption{Comparision between the dust continuum in the simulation (left panel) and observation (right panel) for GM Aur disk from \citet{HuangEtal2020}, both of them are at 1.3 mm, in ALMA Band 6. The intensities are normalized to the maximum values. We indicated the synthesized beam (lower left corner) and the physical scale of the disk (lower right corner). The simulated dust continuum uses the same dust surface density distribution as in \fg{GMaurprofile}, with the used disk parameters labeled on the top left in the panel. The thin white dashed lines indicate the orbit of the imposed planets in the simulation. We overplot the orbits of the hypothesized four planets on top of the observed continuum. }
    \label{fig:gmaur}
\end{figure*}

\section{Two groups of planet-hosting transition disks}
\label{sec:alma}
In the previous section, we show that pebble accretion is able to create the appearance of an inner dust cavity without pressure bumps. Our scenario is consistent with the disks with deep dust cavities but no apparent deep gas cavities, such as GM Aur \citep{HuangEtal2020}. This contrasts with disks hosting giant planets, such as PDS 70 \citep{KepplerEtal2018}, where a gas giant opens up a deep gap in the gas disk, trapping dust in the pressure maximum outside the gas gap.

We propose that there are two groups of \textbf{planet-hosting} transition disks based on the planet's mass:
\begin{itemize}
    \item Above pebble isolation mass: In this group, the planet(s) within the disk exceeds the local pebble isolation mass. The presence of gas giant(s) disrupts the gas disk, leading to pressure bumps that affect the dust profile.
    \item Below pebble isolation mass: Here, the planet(s) within the disk is below the local pebble isolation mass. The gas disk remains undisturbed by the planet(s), while the pebble flux is reduced at planet orbit(s) due to pebble accretion.
\end{itemize}
We select two example sources, representing the two groups respectively, and compare the simulated continuum with observations. The sources are GM Aur and PDS 70. 
The source properties and the information of the used observational data are shown in \tb{source}.

The continuum image is generated using \texttt{RADMC3D} \citep{DullemondEtal2012}. We input 3D dust volume density distribution into \texttt{RADMC3D}. In the vertical direction, the dust density is assumed to follow Gaussian distribution with the scale height calculated via \eq{scaleheight}. The 1D dust surface density distribution can then be projected to 3D volume density distribution. The \texttt{DSHARP} dust opacity \citep{BirnstielEtal2018} is used and is calculated via \texttt{optool} \citep{DominikEtal2021}. The dust follows the default power law distribution with a minimum size of 0.05 $\mathrm{\mu m}$, a maximum size of 3 $\mathrm{mm}$, and a power law slope of 3.5. 



\subsection{Below pebble isolation mass: GM Aur}
\label{sec:gmaur}
GM Aur is of particular interest for this paper as it is a transition disk with distinctive features in its continuum. It exhibits two rings, with the inner ring displaying an outer "shoulder" around $R\approx40$ au. Notably, the inner part of this ring appears steeper than the outer part. This characteristic draws our interest because rings induced by pebble accretion often exhibit similar features -- sharp inner edge but smooth outer edge. In addition, \cite{ZhangEtal2021} ruled out a deep gas cavity in GM Aur using ALMA observations of CO isotopologues.

\begin{figure*}
    \centering
    \includegraphics[trim={0 0 2.7cm 0},clip, height=6cm]{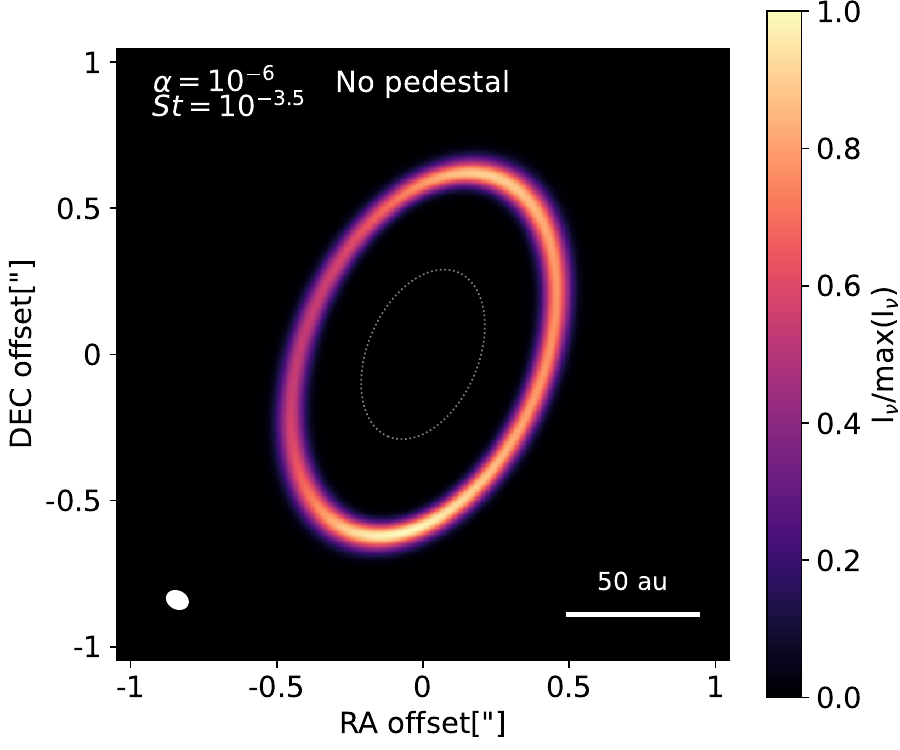}
    \includegraphics[trim={1.8cm 0 2.7cm 0},clip, height=6cm]{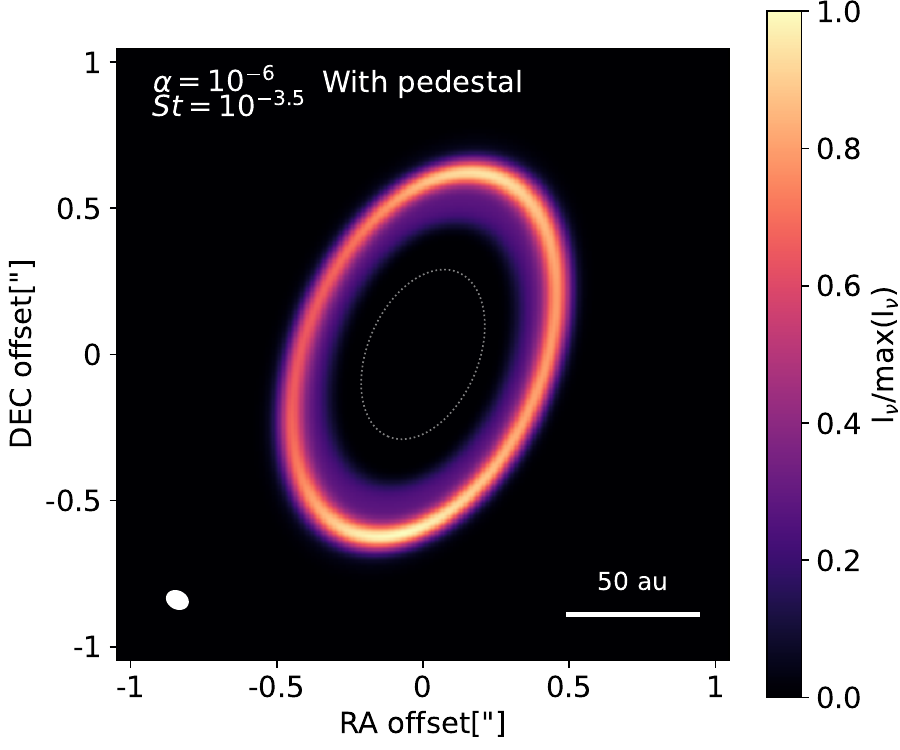}
    \includegraphics[trim={1.8cm 0 0 0},clip, height=6cm]{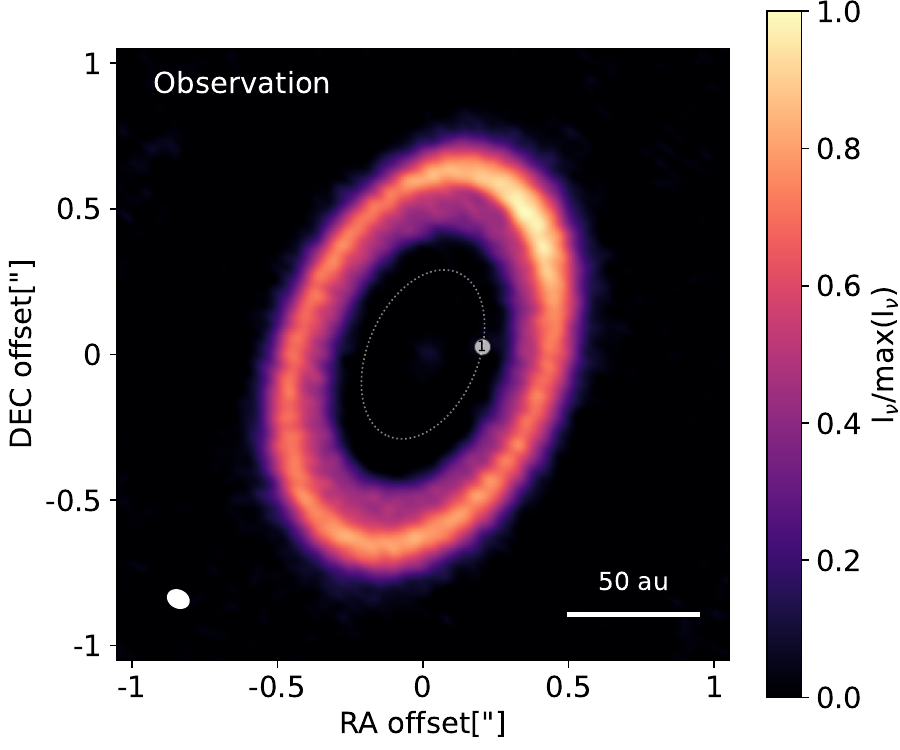}
    \caption{Comparison between the dust continuum in the simulation (left and middle panels) and observation (right panel) for PDS 70 disk from \citet{KepplerEtal2018}, all of them are at 0.9 mm, in ALMA Band 7. The intensities are normalized to the maximum values. We indicated the synthesized beam (lower left corner) and the physical scale of the disk (lower right corner). In the left panel, the dust profile is from the dust evolution simulation. In the middle panel, we add a small dust population onto the simulated dust profile, to simulate the inner pedestal inside the ring.  }
    \label{fig:pds70}
\end{figure*}

For these reasons, we tailor the planet and disk properties in the feasible range following \tb{parameter}, to show the physical consistency between GM Aur and our proposed pebble accretion scenario. Finally, we include four planets, with the inner three planets in a 3:2 resonance chain while the outermost planet does not join. The location of the outermost planet is set to match the position of the observed outer ring in GM Aur. 
The planets are at {20.0, 26.2, 34.3, 75.5 au} orbits, respectively. The planet masses are no more massive than half of the local pebble isolation mass, they are {16.1, 19.8, 24.4, 30.2} $M_\oplus$, respectively. The viscosity $\alpha$ is chosen to be $10^{-6}$ and $\mathrm{St}$ is $10^{-3.5}$. 

The resulting dust surface density profile is shown in \fg{GMaurprofile}. At each planet's orbit, the dust profile shows a cliff-like feature. The inner three planets are half of the local pebble isolation mass. They are responsible for the inner dust cavity, as they accrete the drifting dust efficiently. The efficient accretion is consistent with \fg{dust_profile}. The fourth planet is about 1/3 of the local pebble isolation mass. {The mass of planet e is chosen such that the inner ring is as bright as in the ALMA continuum image \citep{HuangEtal2020}. }

In addition, we find that the planet orbits always coincide with the inner edge of the rings. This is because the inner edge of the rings is the location where the dust surface density drops, which is caused by pebble-accreting planets in our scenario. This is different from the gap opening scenario, where the planets are usually located at the center of the gap \cite[see e.g.][]{KanagawaEtal2017}.

The corresponding continuum image for the simulated dust profile is displayed in \fg{gmaur}, in comparison with the observed image of GM Aur \citep{HuangEtal2020}. The orbits of the imposed planet are indicated by the dashed lines and overplotted on top of the observed continuum image. The longitudes of the planets on their orbits are consistent with the scenario that they are connected by a Laplace resonance chain. 

After convolving the image with the beam, the two rings are still observed in the simulated continuum which is consistent with the observation. 
As expected, the inner ring in the simulated GM Aur continuum has an {extended outer} edge than the inner, which is consistent with the observations. {ALMA also shows that there is an extended outer dust disk of the second ring \citep{HuangEtal2020}, indicating the ongoing dust drifting process.}

{The gap between the two rings is not as pronounced as observed. One possible explanation is that planet e has grown beyond the gap-opening mass. The pressure bump created by planet e prevents the inward drift of small pebbles, which could account for the deeper gap between the two rings seen in observations. If this is the case, planet e must have reached this size within the last few $\sim$ kyr; otherwise, most of the dust would have been trapped in the outer ring, making the inner ring significantly fainter due to dust radial drift. }

{The inner edge of the simulated ring is not extremely sharp because a small percentage of the pebble flux bypasses the orbit of planet d without being accreted. Observations show a similar feature, though with a shallower gradient. Increasing the mass or the eccentricity of planet d would result in a sharper inner edge that is more consistent with observation. }

{We include three planets within the cavity to ensure that most of the dust pebble flux is accreted. In \se{gmaur3}, we also explore the scenario where planet b is removed, but find that the resulting cavity is not as deep as observed. Enhancing the accretion efficiency of pebble flux and clearing the cavity can be achieved by increasing the number of planets, increasing planet mass and putting planets on eccentric orbits \citep{LiuOrmel2018}. }

{We managed to explain the clear inner cavity along with dust drifting outer disk in GM Aur by a sequence of pebble accreting planets.} A detailed fitting going through larger parameter space could still improve the match to the observed continuum image in the future, and potentially give a firmer prediction of the properties of planet candidates. 

\subsection{Above pebble isolation mass: PDS 70}
\label{sec:pds70}
PDS 70 is known to host two directly observed giant planets that create a large gap in both the gas and dust disks \citep{HaffertEtal2019}. This system provides valuable insights into disk structures with massive giant planets. Therefore, we select PDS 70 as a representative disk hosting giant planets with masses above the pebble isolation mass. 

To simulate the outer ring in PDS 70, we impose a 10 $M_\mathrm{J}$ mass planet at 34.3 au. We adopt the analytical shape of the gas gap opened by the planet following \cite{KanagawaEtal2017} and \cite{PichierriEtal2024ii}. The detail is illustrated in \se{gapopen}. The dust diffusion and radial drift are considered during its transport. The dust ring structure does not sensitively depend on the disk parameter we choose. As an example, we use $\alpha=10^{-6}$ and $\mathrm{St}=10^{-3.5}$, which is same as the values taken in \se{gmaur}.

The simulated dust profile is used to generate a continuum image through \texttt{RADMC3D} and presented in the left panel in \fg{pds70}. The dust drifts inward and accumulates at the pressure maximum outside the gap opened by the gas giant. 
The observed Band 7 high-resolution continuum image \citep{KepplerEtal2018} is shown in the right-hand panel. The simulated dust ring is located at the separation similar to the observation.

An apparent discrepancy between observation and simulation is that the observed continuum shows a pedestal at the inner edge of the dust ring while the simulated dust ring is radially symmetric. 
We note that including small grains might be able to explain the pedestal \citep{PinillaEtal2024i}. 
We think that there are a couple of possible scenarios to explain the pedestal. The pedestal may be composed of i) small dust grains following the gas gap profile as they are too small to get trapped in the pressure bump; ii) small dust entrained by the accretion flow onto the planet \citep{TociEtal2020}; iii) small dust entrained by the irradiated flow from the inner gas wall \citep{FranzEtal2022}.

\begin{table*}[!ht]
    \centering
    \caption{Sample of transition disks with dust ring ratios, CO properties and stellar properties.}
    \label{tab:classification}    
    \begin{tabular}{l|llc|cccl|c|lll}
    \hline
Target	&	$R_1$	&	$R_2$	&	Ped.$^a$	&	CO	&	$^{12}$CO &	$^{13}$CO 	&	Ref.	&	Class	&	$d$	&	$M_*$	&	$\log M_{acc}$	\\
	&		&		&		&	analysis&	gap?	&	gap?		&		&		&	(pc)	&	($M_{\odot}$)	&	(M$_{\odot}$ yr$^{-1}$)	\\
 \hline
PDS70	&	0.7$\pm$0.1	&	-	&	Y	&	deep	&	Y	&	Y	&	1,2	&	A	&	113	&	0.8	&	-10.2	\\
SR21	&	0.7$\pm$0.1	&	-	&	M	&	deep	&	Y	&	Y	&	3,4	&	A	&	138	&	2.0	&	-7.9	\\
HD169142	&	0.8$\pm$0.2	&	-	&	Y	&	deep	&	Y	&	Y	&	5	&	A	&	114	&	2.0	&	-8.7	\\
HPCha	&	0.8$\pm$0.1	&	-	&	N	&	-	&	-	&	-	&	-	&	A	&	160	&	1.4	&	-9.0	\\
LkCa15	&	0.4$\pm$0.2	&	1.2$\pm$0.1	&	Y	&	moderate	&	M	&	M	&	5	&	A/B?	&	158	&	0.8	&	-8.4	\\
HD135344B	&	0.9$\pm$0.2	&	0.7$\pm$0.1	&	Y	&	deep	&	Y	&	Y	&	3,6	&	A	&	136	&	1.6	&	-7.4	\\
DoAr44	&	1.0$\pm$0.2	&	0.9$\pm$0.1	&	Y	&	deep	&	-	&	Y	&	3	&	A	&	146	&	1.0	&	-8.2	\\
J16042165	&	1.1$\pm$0.3	&	0.8$\pm$0.1	&	N	&	deep	&	Y	&	Y	&	7,8	&	A	&	150	&	1.2	&	-10.5	\\
ISO-Oph2	&	1.1$\pm$0.1	&	-	&	M	&	-	&	N	&	-	&	9	&	A/B?	&	144	&	0.5	&	-8.5	\\
RXJ1633.9	&	1.1$\pm$0.1	&	-	&	M	&	-	&	Y	&	-	&	9	&	A/B?	&	141	&	0.8	&	-10.0	\\
HD97048	&	1.2$\pm$0.1	&	1.2$\pm$0.1	&	N	&	-	&	Y	&	Y	&	10,11	&	A	&	185	&	2.4	&	-8.2	\\
SYCha	&	1.3$\pm$0.0	&	-	&	N	&	-	&	N	&	-	&	12	&	B	&	183	&	0.8	&	-9.4	\\
HD100453	&	1.4$\pm$0.2	&	-	&	N	&	-	&	N	&	-	&	13	&	B	&	104	&	1.5	&	<-8.3	\\
HD100546	&	1.4$\pm$0.2	&	-	&	N	&	deep	&	Y	&	Y	&	14	&	A/B?	&	110	&	2.2	&	-7.0	\\
SR24S	&	1.6$\pm$0.2	&	1.3$\pm$0.1	&	N	&	-	&	N	&	N	&	9,15	&	B	&	114	&	1.5	&	-7.2	\\
GMAur	&	1.5$\pm$0.2	&	1.4$\pm$0.2	&	N	&	moderate	&	N	&	M	&	16	&	B	&	160	&	1.0	&	-8.3	\\
J16100501	&	1.8$\pm$0.3	&	2.2$\pm$0.8	&	N	&	-	&	-	&	-	&	-	&	B	&	145	&	0.6 &	-8.6	\\
CSCha	&	2.0$\pm$0.5	&	-	&	N	&	-	&	Y	&	-	&	17	&	B	&	176	&	1.5	&	-8.3	\\
WSB82	&	3.0$\pm$0.2	&	-	&	N	&	-	&	N	&	-	&	9	&	B	&	156	&	-	&	-	\\
\hline
    \end{tabular}\\
    $^a$ Presence of a pedestal: Y=Yes, N=No, M=Maybe.\\
Refs. 1) \citet{MuleyEtal2019}, 2) \citet{FacchiniEtal2021}, 3) \citet{vanderMarel2016}, 4) \citet{Yang2023}, 5) \citet{LeemkerEtal2022}, 6) \citet{Casassus2021}, 7) \citet{Dong2017} 8) \citet{Stadler2023}, 9) \citet{Antilen2023}, 10) \citet{Pinte2019}, 11) \citet{Woelfer2023}, 12) \citet{Orihara2023}, 13) \citet{Rosotti2020}, 14) \citet{LeemkerEtal2024i}, 15) \citet{Pinilla2016}, 16) \citet{Zhang2021}, 17) \citet{Kurtovic2022}.
\end{table*}

\begin{figure*}
    \sidecaption
    \includegraphics[trim={0 0.4cm 0 0.3cm},clip, width=1.1\columnwidth]{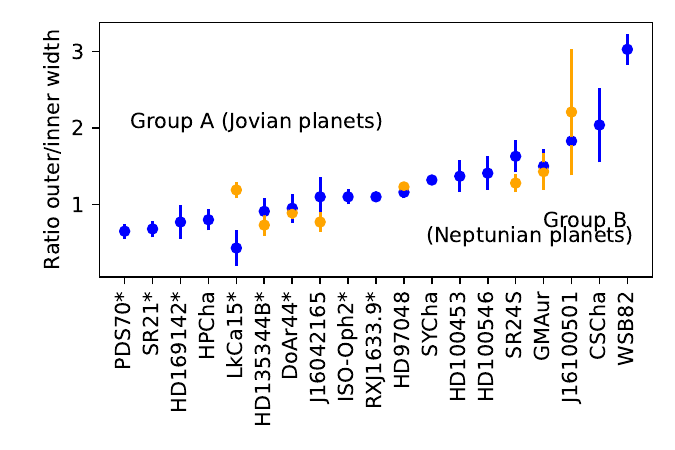}
    \caption{Ratio between outer and inner width of the dust rings in a sample of transition disks, as derived in \se{gaussfit}. The blue symbols correspond to the inner ring, and the orange symbols to the outer ring, if present. Disks with an inner pedestal have been marked with a star. The targets are sorted by the average value of the ratio. Lower values $\lesssim$1 are thought to correspond to Group A (Jovian protoplanets with a dust trap), and higher values $\gtrsim$1 are thought to correspond to Group B (Neptunian protoplanets which are creating an apparent cavity by pebble accretion. 
    }
    \label{fig:classification}
\end{figure*}
While explaining the pedestal is out of the scope of this work, we here simply assume there is a population of small dust grains that are well-coupled with gas. The dust distributes in the outer slope in the gas gap with a surface density of one percent of the gas:
\begin{equation}
\begin{aligned}
\Sigma_{\mathrm{d,sm}}(r)
     & = \left\{\begin{array}{ll}
        0.01\Sigma_\mathrm{g}(r) &\mathrm{for} \, \Delta r_1<(r-r_\mathrm{p})<\Delta r_2, \\
        0\,\mathrm{g/cm^2}  & \mathrm{otherwise}.
    \end{array}
    \right.
\end{aligned}
\end{equation}
The planet's orbit is $r_\mathrm{p}$, $\Delta r_1$ is the gap bottom radius and $\Delta r_2$ is the gap edge radius. The simulated dust continuum image after adding the small dust population is shown in \fg{pds70} middle panel. By construction, we now observe the inner pedestal within the main ring in the generated continuum. Whether the protoplanetary disks with gas giants always exhibit rings with pedestals outside the planets' orbits needs to be examined in the future, when more proto-giant planets are confirmed. 

Regardless of the pedestal, the dust ring morphology outside a massive planet is distinct from a dust ring outside a sequence of pebble-accreting planets. Compared to the dust rings induced by pebble-accreting low-mass planets (e.g., in GM Aur), the dust rings trapped in the pressure maximum caused by gap-opening planets have more radially symmetric rings (maybe show even smoother inner edges in observation e.g., in PDS 70). This difference arises from the physics involved: in PDS 70, the diffusion of the dust near the gas pressure maximum results in a symmetric ring. Conversely, in GM Aur, the sharp cutoff of pebble flux by the planet's pebble accretion creates a cliff-like feature at the ring's inner edge.

\begin{figure*}
    \centering
    \includegraphics[width=1.7\columnwidth]{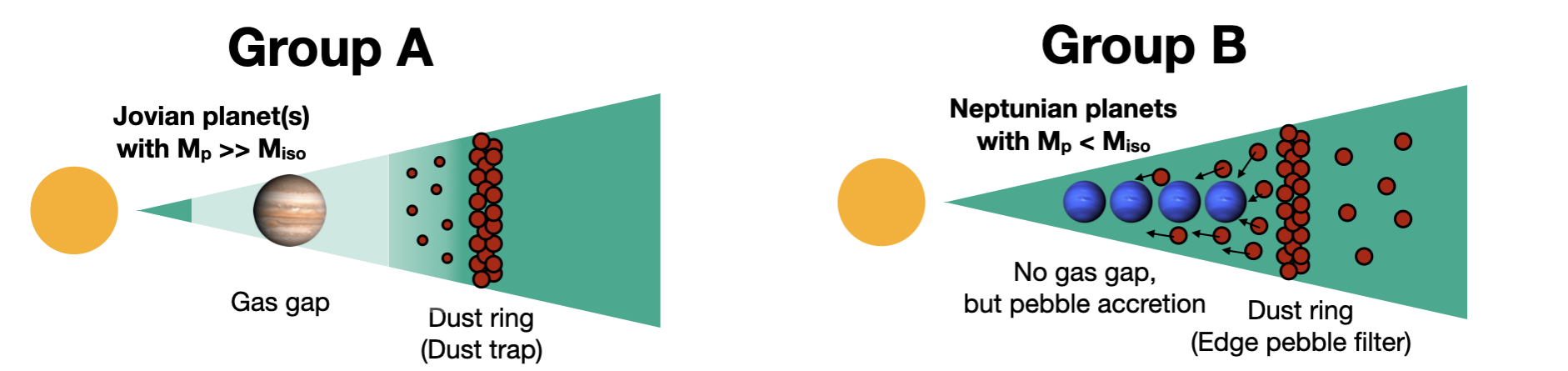}
    \caption{Cartoon demonstrating the two proposed mechanisms responsible for the transition disk cavity appearance. For disks classified as Group A it is thought that one or more Jupiter-like planets with a mass well above the pebble isolation mass have carved a gas gap, trapping the millimeter dust in a narrow ring outside its orbit, whereas for disks classified as Group B a number of Neptunian planets with a mass below pebble isolation mass are still actively accreting pebbles, thereby creating the appearance of a dust cavity as the inner edge of the dust ring is acting as a 'filter' where pebbles are accreted, and the pebbles in the dust ring are not trapped, but follow a smooth dust density pattern.}
    \label{fig:transition_sketch}
\end{figure*}

\begin{figure*}
    \sidecaption
    \includegraphics[trim={0 0.4cm 0 0.8cm},clip, width=\columnwidth]{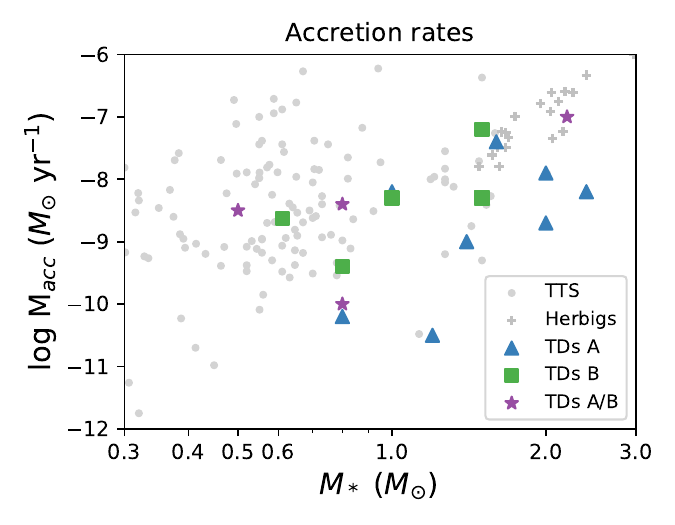}
    \caption{Accretion rates of the transition disks in our sample (colored symbols) compared with those of the larger disk population, taken from \citet{Manara2023, Wichittanakom2020}. Group A transition disks appear to have very low accretion rates compared to the general disk population, whereas Group B transition disk accretion rates are more typical. For the Group A/B transition disks the classification from dust and CO observations is still unclear.}
    \label{fig:stellar}
\end{figure*}

\subsection{A sample of transition disks}
\label{sec:statistics}
Whether the planet in the disk is below or above the local pebble isolation mass results in different radial ring morphologies. Motivated by the apparent morphologies of PDS70 and GM Aur, we calculate the ratio of the outer to inner width of dust rings ($R=\sigma_\mathrm{outer}/\sigma_\mathrm{inner}$) in the radial profiles of a larger sample of $\sim$20 transition disks (see \se{gaussfit}). The results are shown in \fg{classification}. If two rings are present, both ring ratios are shown. We tentatively classify the transition disks into two groups according to the dust ring structure:
\begin{itemize}
    \item \underline{Group A}: The width ratio of the ring is $\lesssim$1 and a pedestal may be present.
    \item \underline{Group B}: The width ratio of the ring is $\gtrsim$1 and no pedestal is present.   
\end{itemize}

Group A is consistent with giant planet disks with the planet mass above the pebble isolation mass. Group B is consistent with the disks with the planet mass below the pebble isolation mass. The difference is illustrated in \fg{transition_sketch}. 

Furthermore, we search the literature for information about the gas inside the cavity through spatially resolved CO 3-2 or 2-1 isotopologue ALMA images. This helps distinguish between cavities with a protoplanet massive enough to carve a deep gas gap (Group A) and cavities with less massive protoplanets that cannot carve such a gap, like GM Aur (Group B). Only CO data with sufficient spatial resolution to fully resolve the dust cavity by at least 3 beams are considered. For half of our sample, physical-chemical modeling has been used to derive the gas cavity depth \citep[e.g.][]{vanderMarel2016}, which is often found to be at least 2 orders of magnitude depleted in density ('deep'), but sometimes just a factor of a few ('moderate'). Deep gas cavities usually show a deficit of $^{12}$CO emission in the center of the disk, as even the $^{12}$CO emission becomes optically thin. 
For transition disks without physical-chemical analysis, it is still possible to assess the possibility of a deep gas cavity considering the presence of a $^{12}$CO and/or $^{13}$CO deficit. \Tb{classification} summarizes the results for all targets in our sample for both the dust ring ratios and the CO properties as well as its stellar properties \citep[taken from][]{vanderMarel2023}. Based on CO and dust ring properties, each disk is then classified as either A-type or B-type. Some disks have inconsistencies when considering the dust profile and the CO properties, or are on the borderline: these are marked as A/B.

Using the classifications derived in \tb{classification}, we finally explore the accretion rates of the two groups. \Fg{stellar} shows the accretion rates of the transition disks in our sample compared with the accretion rates of all nearby T Tauri stars, using the stellar data from \citet{Manara2023} as well as the accretion rates of all nearby Herbig stars, using the stellar data from \citet{Wichittanakom2020}.

When comparing the accretion rates of the two groups, Group A transition disks appear to have very low accretion rates compared to the general disk population, whereas Group B transition disk accretion rates are 1-2 orders of magnitude higher and more similar to typical accretion rates. This is consistent with a scenario where the protoplanets in Group A transition disks are almost finished accreting gas, as they have already grown to (super-)Jupiter sizes and cleared a gap in their gas distribution, whereas Group B transition disks contain protoplanets that have not started runaway gas accretion yet, which is expected for planets well below pebble isolation mass with high pebble accretion rates \citep{BrouwersEtal2021, OrmelEtal2021}. 

\section{Discussion}
\label{sec:discussion}
\subsection{Model assessment}
\label{sec:assessment}
We have simulated dust transport, accounting for the reduction of pebble flux due to pebble accretion. This approach enables us to analyze the resulting dust surface density profile across various disks hosting low-mass planets (below pebble isolation mass). The planets are assumed to have zero eccentricities, while eccentric planets can increase the pebble accretion efficiency by 3-5 \citep{LiuOrmel2018}. 


We neglect the perturbation that the planet exerts on the gas disk. When a planet approaches the pebble isolation mass, it can still create shallow gaps. \cite{BitschEtal2018} suggest that a planet at this mass can generate a gap with a depth of about 10-20\%. Dust may accumulate at such shallow pressure maxima. Additionally, gas polar inflow and radial outflow induced by the planet's gravity can inhibit pebble drift \citep[][and references therein]{KuwaharaEtal2022, BiLin2024i}, and possibly decrease pebble accretion efficiency \citep{KuwaharaKurokawa2020, OkamuraKobayashi2021}. 

This work focuses on capturing the changes in pebble flux and surface density profiles due to pebble accretion
. Arguably, our model applies even to the planets with their mass slightly above the pebble isolation mass because the timescale for gap opening correlates with the viscous timescale. Therefore, when a planet doubles its mass, the gap-opening process may not be complete \citep{Bergez-CasalouEtal2020}, allowing dust to drift inward while the pressure bump is forming.

The model is not very sensitive to the gas temperature profile. The dust scale height and pebble isolation mass increase with the temperature increase. Increasing scale height decreases the probability for the pebbles to hit one planet, which is lower in the 3D limit. In contrast, larger pebble isolation mass makes the planets more difficult to open the gap, which enhances pebble accretion efficiency because of the increase of the cross-section between the larger planet and the pebbles. 


We do not account for the radial diffusion of dust when studying the resulting dust profile after pebble accretion. The inclusion of radial diffusion may have two main differences: On one hand, dust diffusion could slow down the dust radial drift speed \citep{GerosaEtal2024i}. The random motion of dust particles could increase their chances of close encounters with the planet and potentially enhance the pebble accretion rate. On the other hand, if the dust is too diffusive, the dust may bypass the planet’s orbit rather than being accreted. {We assess that dust diffusion during pebble accretion is not important in \se{diffusion}, where we show that the distance dust particles diffuse is smaller than planet's Hill radius. However, we acknowledge that the full impact of radial dust diffusion on pebble accretion requires more detailed investigation.}


\subsection{Feasibility of the "sweet spot" on the accretion map}
\label{sec:sweetspot}
Low viscosity ($\alpha=10^{-5}-10^{-6}$) is justified both theoretically and observationally. High stellar accretion rates observed in protoplanetary disks have been interpreted as evidence for $\alpha=10^{-3}-10^{-2}$ in the classic viscous disk model \citep{Lynden-BellKalnajs1972}, where viscosity may arise from magneto-rotational instability \citep{BalbusHawley1991}. However, in regions with low ionization and weak coupling between gas and magnetic fields, hydrodynamic or non-ideal MHD effects could dominate angular momentum transfer, resulting in stellar accretion rates matching observations \citep{BaiStone2013, FlockEtal2020}. In such scenarios, turbulent parameters can be as low as $\alpha=10^{-4}-10^{-6}$, as preferred in our study. Observations of inclined protoplanetary disks, such as HL Tau and Oph 163131, reveal settled dust profiles in the midplane, indicating low turbulence with $\alpha\lesssim10^{-5}$ \citep{PinteEtal2016, VillenaveEtal2022}. From line broadening, \cite{FlahertyEtal2020} suggests weak turbulence for MWC 480 and V4046 Sgr.

Recent simulations of dust size evolution suggest a near-monodisperse distribution centered around $\mathrm{St}\sim10^{-3}$ \citep{DominikDullemond2024}. The maximum size of dust particles is constrained by fragmentation, radial drift, and bouncing \citep{KellingEtal2014, KrussEtal2016}, limiting the dust Stokes number to $\sim10^{-3}$ \citep{DominikDullemond2024}. In low $\alpha$ disks, the bouncing barrier primarily restricts the upper limit of the dust size distribution, resulting in near-monodisperse size distributions in simulations. Notably, the Stokes number depends on the gas surface density, $\mathrm{St}\sim s/\Sigma_\mathrm{g}$. Our Group B disks are likely in an early phase with relatively massive gas disks, so the corresponding grain Stokes number (with the same size $s$) could be relatively small compared to other more evolved disks.

\subsection{Emerging the planet formation story in transition disks}
Since the direct detection of proto-giant planets in the PDS 70 disk, (super-)Jovian-mass planets have been favored for explaining transition disks. Their substantial mass allows them to strongly perturb the disk and create cavities or gaps \citep[e.g.,][and references therein]{DongEtal2015, DongEtal2017, LiEtal2019, Garrido-DeutelmoserEtal2023i, WuEtal2023}.
In this study, we propose that a chain of planets below the pebble isolation mass could also account for transition disks, named Group B disks. Group A disks are those hosting (super-)Jupiters.  

Group A transition disk may evolve from B transition disks. If Jovian-mass planets form through core accretion \citep[e.g.,][]{PerriCameron1974}, they must first pass through a stage where they are of Neptunian size.{ Our study shows that dust cavities begin to appear even when planets are still low-mass and the gas disk remains largely unperturbed. As planets grow, while pressure bumps form gradually, dust continues to drift and accumulate at these pressure bumps. Over time, we expect the dust ring morphology to become more radially symmetric (possibly even developing an inner pedestal, as seen in PDS 70, \se{pds70}) and the Group B disk to gradually transition into a Group A disk. }
However, Jovian-mass planets may also form directly via Gravitational Instability \citep[GI,][]{Boss1997}. In such cases, the period of maintaining Neptunian mass may be too short for dust to drift. Potentially, the Group B stage is skipped in the case of GI and the two groups do not necessarily represent evolutionary stages. 
It is worth noting that GI may also form super-Earth to Neptune mass planets \citep{KubliEtal2023}, but this still needs further investigation.

Group B disks also do not necessarily evolve into Group A from the perspective of gas accretion. The growth of Neptunian-mass planets via pebble accretion \citep{OrmelLiu2018, JohansenBitsch2019} and subsequent gas accretion \citep{IdaEtal2018} can lead to Jovian-mass planets. Gas accretion is a complex process. It is initially governed by Kelvin-Helmholtz contraction but then limited by the amount of gas entering the Hill-sphere of the planet \citep{TanigawaTanaka2016}. The envelope contraction speed depends on the cooling timescale. If the atmosphere opacity is high, or the atmospheric recycling is efficient \citep[e.g.][]{OrmelEtal2015}, the gas accretion can be prolonged. If the gas disk disperses before the runaway gas accretion on the planet can happen, the Group B transition disk can never evolve to A. 

In a Group B transition disk, the creation of a large dust cavity in the protoplanetary disk necessitates a multi-planet system, typically comprising 3 or more planets. Such large planet multiplicity aligns with observations of certain systems. For instance, the four gas giants in the HR 8799 system \citep{GozdziewskiMigaszewski2020} may grow from a sequence of low-mass planets. Similarly, the PDS 70 system has two confirmed planets and possibly a third one \citep{ChristiaensEtal2024i}.
However, it's noteworthy that the observed multiplicity of planets at wide orbits (beyond $\gtrsim$10 au) might be lower than expected. Several factors contribute to this trend. 1) The innermost planet may not have yet encountered the migration barrier. Planets migrate inward to the region $<10$ au in Type I/II regime \citep{PaardekooperEtal2010, KanagawaEtal2018}. 2) Planets may undergo inefficient growth, and the disk may dissipate before they enter the runaway growth phase, analogous to the giant planets in our solar system \citep{GurrutxagaEtal2024}. Small planets on wide orbits are challenging to detect. 3) As the number of planets in a system increases, the dynamical stability of the system tends to decrease \citep{PichierriMorbidelli2020}. Consequently, some planets may become inclined/unbound from their host stars following the onset of instability, so a multi-Neptunian protoplanet system may end up with a smaller number of bound Jovian planets.

\subsection{Consequence for planet detection}
The planets in Group B disks are often Neptune-sized and located at the inner edge of the dust rings. Current instruments are not sensitive enough to detect protoplanets below a few Jupiter masses \citep{Asensio-TorresEtal2021, ChoksiChiang2024}. Even with more sensitive instruments in the future, detecting Neptunian protoplanets would still be extremely challenging due to their spatial proximity to the dust ring. Selecting the right targets (with a Group A morphology) would thus be the best strategy for future direct imaging campaigns.


\section{Conclusions}
\label{sec:conclusion}
We use dust transport models to study the dust surface density profiles in pebble-accreting multi-planet systems, in order to understand how the dust profile in protoplanetary disk changes with planet mass, disk viscosity, and dust size. We then compare the dust substructure morphology induced by pebble-accreting planets with those caused by pressure bumps. These are our main findings:
\begin{enumerate}
    \item Multi-Neptunian planets (below the local pebble isolation mass) can sculpt deep cavities in the dust disk, potentially contributing to the appearence of some observed transition disks. 
    \item Planet-hosting transition disks may consist of two sub-groups. In Group A disks, gas giants open deep gas gaps and truncate the disk. In Group B disks, a sequence of Neptune-mass planets accrete all the inward pebble flux, leaving the gas disk unperturbed.
    \item Rings formed by pebble-accreting planets often feature sharp inner edges and smooth outer shoulders (Group B disks), unlike the more symmetric rings induced by pressure bumps exterior to the gas gaps carved by massive gas giants (Group A disks).
    \item In Group B disks, planets harbor near the inner edge of the induced rings, whereas in Group A disks the giant planets are located at the center of gaps/cavities.
    \item Our classification of transition disks (Group A and B disks) is supported by observations. Transition disks with rings featuring {extended outer but} sharp inner edges coincide with higher stellar accretion rates and lack deep gas cavities inside the dust cavity, likely hosting pebble-accreting planets. The other disks, characterized by smooth rings, have lower accretion rates and deep gas cavities, likely hosting super-Jupiters.
\end{enumerate}
Pebble accretion  alters the dust profile in planet-forming disks, especially in low-turbulence environments. Although Neptunian-size planets at tens of au orbits remain beyond current observational capabilities, they may yield dust rings with unique morphology in the continuum. Future high-resolution ALMA observations hold promise for expanding the sample size and corroborating our findings with enhanced statistical confidence. 

\begin{acknowledgements}
We thank the anonymous referee for their constructive suggestions and comments. We thank Andrew Sellek, Ayumu Kuwahara, Chris Ormel, Logan Francis, Margot Leemker, Ruobing Dong, and Xinyu Zheng for the beneficial discussions. Software: \texttt{AMUSE} \citep{PortegiesZwartEtal2009, PortegiesZwartEtal2013, PelupessyEtal2013}, \texttt{RADMC3D} \citep{DullemondEtal2012}, \texttt{Optool} \citep{DominikEtal2021}, \texttt{Matplotlib} \citep{Hunter2007, CaswellEtal2021}. 
\end{acknowledgements}

%
%

\bibliographystyle{aa}
\bibliography{ads} 

\begin{appendix}
\section{Numerical treatment of pebble flux reduction}
\label{sec:numerical}
\begin{figure}[ht!]
    \centering
    \resizebox{\hsize}{!}{\includegraphics[width=\columnwidth]{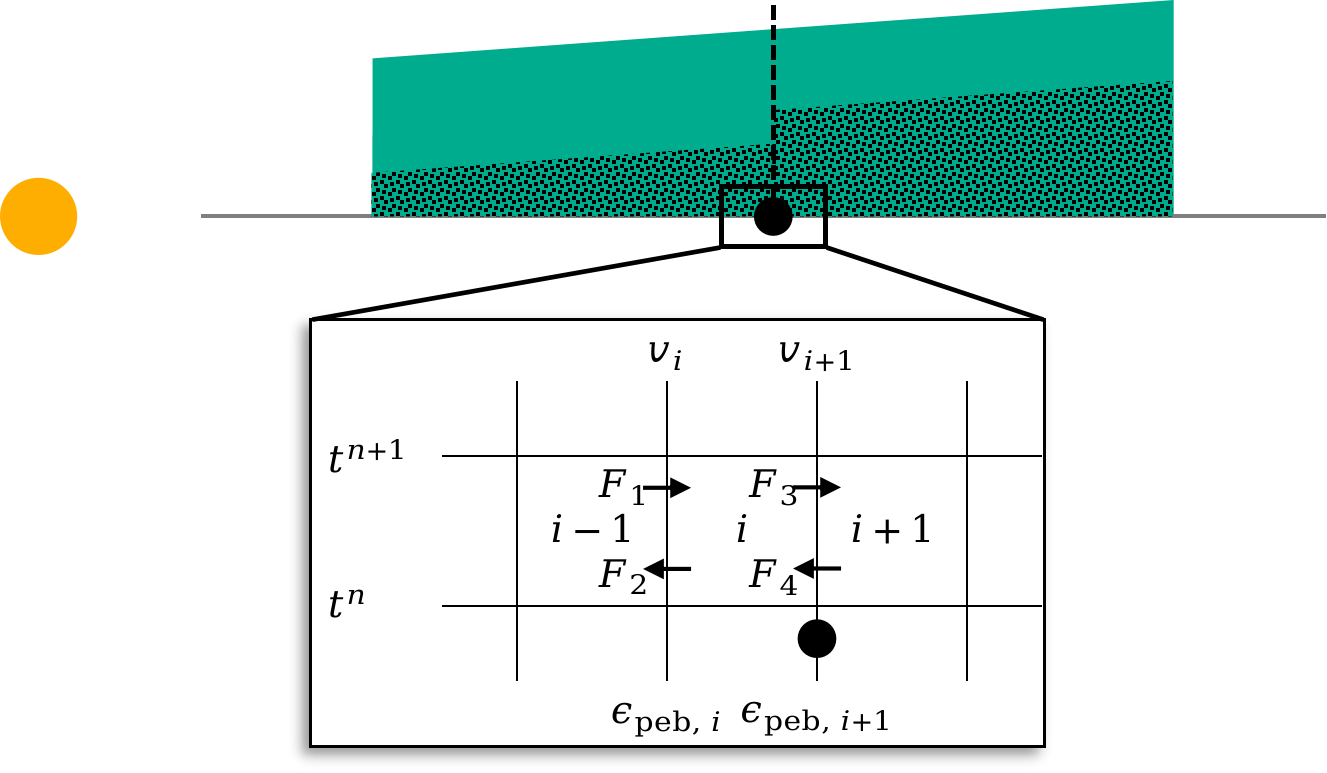}}
    \caption{Sketch of the dust transport calculation on the disk grid $i$, from the time $t^n$ to the time $t^{n+1}$, including pebble accretion. The disk is segmented into $N$ grids. We focus on the region surrounding disk cell $i$ in the zoomed-in view. Arrows depict the direction of various pebble fluxes. The flux going through the planet's orbit $r_i$ is reduced by a factor $\epsilon_{\mathrm{peb},i}$, which is the planet's pebble accretion efficiency. }
    \label{fig:sketch_pebble_accretion}
\end{figure}
To solve for dust transport, we consider a grid of $N$ cells valued by the dust surface density at their centers. An example of calculating the dust surface density in cell $i$ from time $n$ to $n+1$ is a sketch provided in \fg{sketch_pebble_accretion}. the adopted differential form of the advection term is
\begin{equation}
\begin{aligned}
     (\Sigma_{\mathrm{d},i}^{n+1} & -\Sigma_{\mathrm{d},i}^{n}) \frac{\Delta r}{\Delta t}=  
    F_1+F_2+F_3+F_4 ,
\end{aligned}
\end{equation}
where $\Sigma_{\mathrm{d},i}^n$ is the dust surface density in cell $i$ at time $n$, and $F_1$, $F_2$, $F_3$, and $F_4$ are expressed as:
\begin{equation}
\label{eq:F1}
    F_1 = (1-\epsilon_{\mathrm{pebb},\,i}) \cdot \max[v_{i} \Sigma_{\mathrm{d},i-1}, 0],
\end{equation}
\begin{equation}
    F_2 = \min[-v_{i+1} \Sigma_{\mathrm{d},i}, 0],
\end{equation}
\begin{equation}
    F_3 = \min[v_{i} \Sigma_{\mathrm{d},i}, 0],
\end{equation}
\begin{equation}
\label{eq:F4}
    F_4 = (1-\epsilon_{\mathrm{pebb},\,i+1}) \cdot \max[-v_{i+1} \Sigma_{\mathrm{d},i+1}, 0],
\end{equation}
respectively. They represent the pebble flux from cell $i-1$ to $i$ (outward), from cell $i$ to $i-1$ (inward), from cell $i$ to $i+1$ (outward) and from cell $i+1$ to $i$, respectively. The dust velocities at the edges between cells $i-1$, $i$, and $i$, $i+1$ are $v_i$ and $v_{i+1}$. To simulate the pebble accretion effect on the dust, we reduce the incoming flux $F_1$ (\eqb{F1}) and $F_4$ (\eqb{F4}) by factors of $\epsilon_{\mathrm{pebb},\,i}$ and $\epsilon_{\mathrm{pebb},\,i+1}$, which stand for the 3D pebble accretion efficiency on the planets at the two edges. If no planet is included, the corresponding efficiency is set to zero. {For example, in \fg{sketch_pebble_accretion}, a planet is located at the inner boundary of cell $i+1$ but not at $i$, so $\epsilon_\mathrm{pebb,\,i+1}$ is calculated, while $\epsilon_\mathrm{pebb,\,i}$ remains zero. In this case, only $F_4$ is reduced when determining the dust surface density in cell $i$.}

\section{Evaluating small dust diffusion in the context of pebble accretion}
\label{sec:diffusion}
{We evaluate the feasibility of neglecting dust diffusion in our simulations. If dust diffusion is strong, dust particles may travel across the planet's orbit without being accreted. To assess this, we calculate the distance over which dust diffuses during a synodical timescale with the nearby planet and compare it to the planet's Hill radius. }

{The synodical timescale between the planet and dust just outside the planet's Hill radius is given by: }
\begin{equation}
    t_\mathrm{syn}=\frac{4\pi}{3\Omega_\mathrm{K}}\frac{r}{R_\mathrm{Hill}},
\end{equation}
{where $R_\mathrm{Hill}$ is the planet Hill radius and $\Omega_\mathrm{K}$ is the Keplerian frequency at planet's orbit. The diffusivity of small dust particles can be approximated by the gas diffusivity: }
\begin{equation}
    D=\alpha H_\mathrm{g}^2 \Omega_\mathrm{K},
\end{equation}
{Thus, the distance dust diffuses over the synodical period is:}
\begin{equation}
    \Delta r_\mathrm{diff} = \sqrt{Dt_\mathrm{syn}}.
\end{equation}
{If $\Delta r_\mathrm{diff}$ exceeds the planet's Hill radius, the dust is less likely to be accreted due to strong diffusion. Conversely, if $\Delta r_\mathrm{diff}$ is small, the dust can still be accreted. We define the dimensionless quantity $C$ as:}
\begin{equation}
    C=\frac{\Delta r_\mathrm{diff}}{2R_\mathrm{Hill}}=0.1\left(\frac{T}{100\,\mathrm{K}} \frac{r}{10\,\mathrm{au}} \frac{\alpha}{q_\mathrm{p}} \frac{M_\star}{M_\odot}\right)^{0.5},
\end{equation}
{where $q_\mathrm{p}$ is the planet-to-star mass ratio. The dust hardly diffuses across planet orbit if $C$ is smaller than unity. Dust diffusion can be considered negligible if $C$ is smaller than unity. In our simulations, with $\alpha$ ranging from $10^{-4}$ to $10^{-6}$ and the planet about Neptune mass, the value of $C$ remains below unity. The dust diffusion can reasonably be ignored in our pebble accretion simulations. }

\section{Simulated continuum of GM Aur with three planets}
\label{sec:gmaur3}
\begin{figure}[ht!]
    \centering
    \includegraphics[width=\columnwidth]{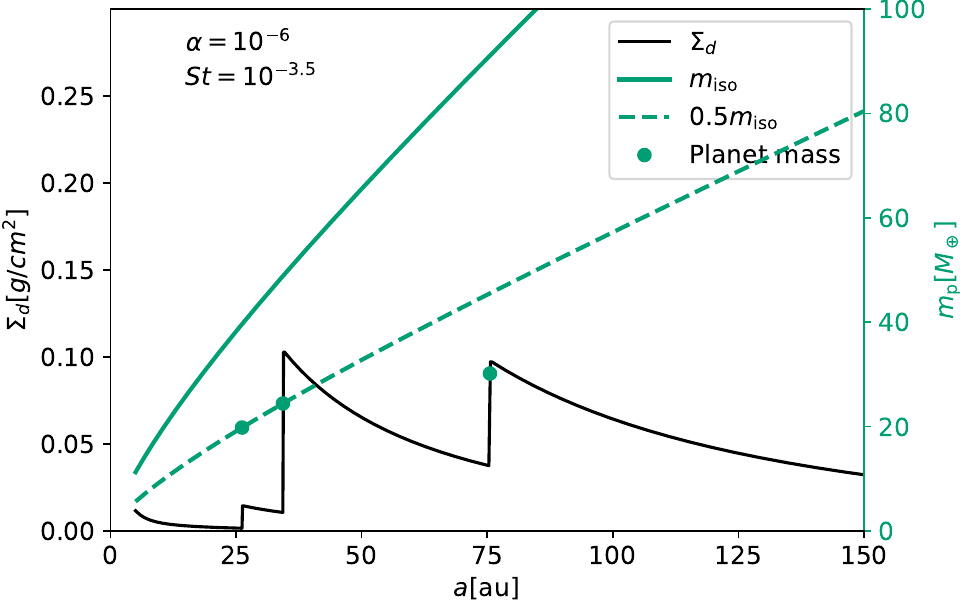}
    \caption{Similar to \fg{GMaurprofile} but with 3 planets. }
    \label{fig:GMaurprofile3}
\end{figure}

\begin{figure*}
    \sidecaption
    {\includegraphics[trim={0 0 2.7cm 0},clip, height=6cm]{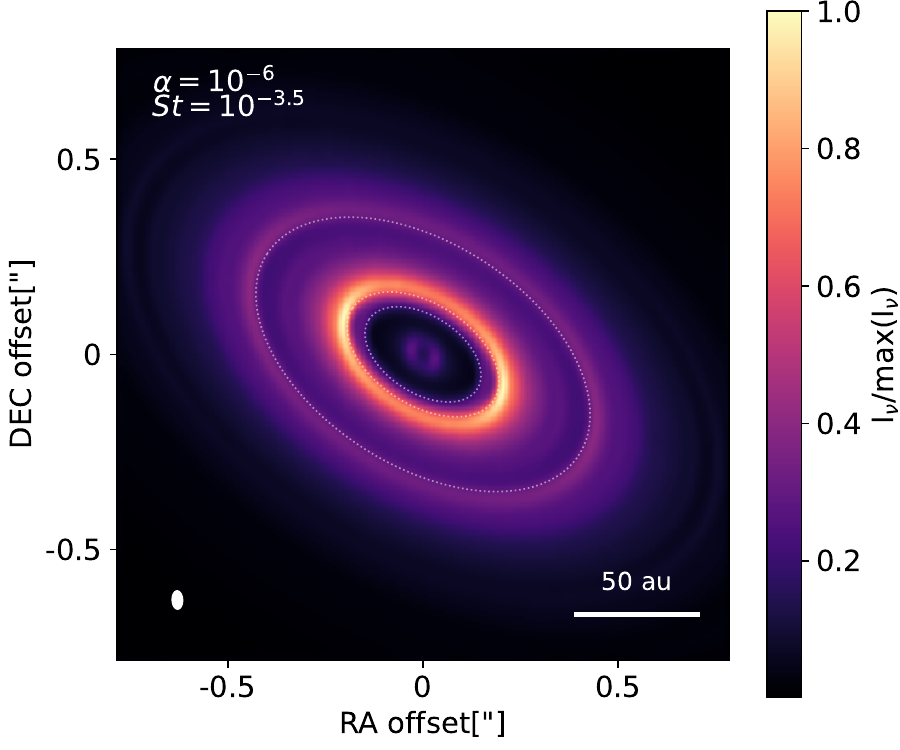}
    \includegraphics[trim={1.8cm 0 0 0},clip, height=6cm]{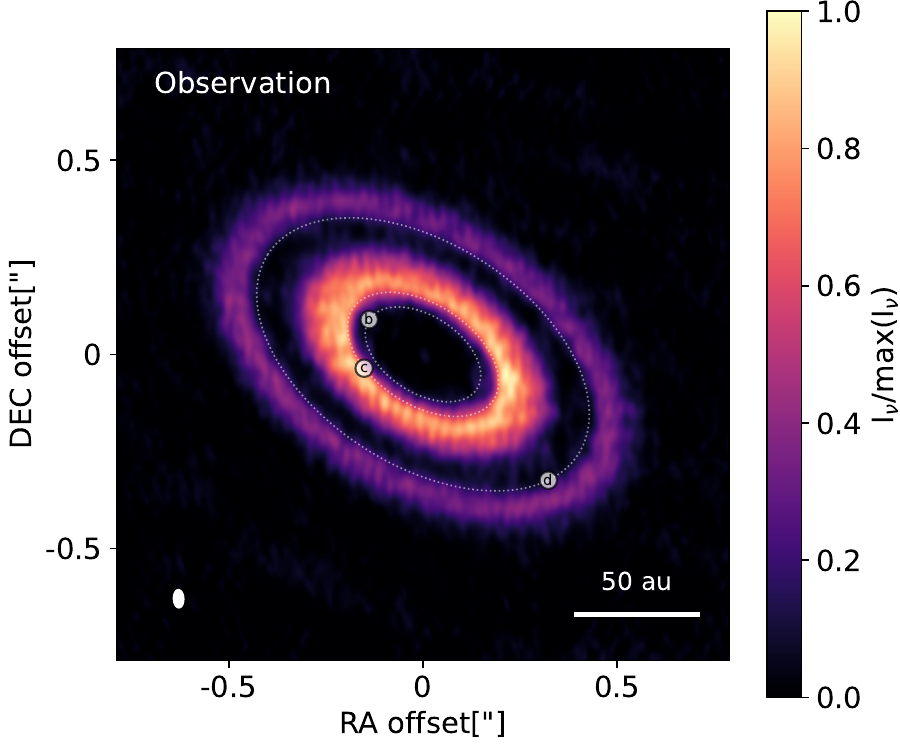}}
    \caption{Similar to \fg{gmaur} but with 3 planets. }
    \label{fig:gmaur3}
\end{figure*}

{Similar to \se{gmaur}, we adopt three planets in the simulation by removing the innermost planet in \se{gmaur}.
The planets are at {26.2, 34.3, 75.5 au} orbits, respectively. The planet masses are no more massive than half of the local pebble isolation mass, they are {19.8, 24.4, 30.2} $M_\oplus$, respectively. The viscosity $\alpha$ is chosen to be $10^{-6}$ and $\mathrm{St}$ is $10^{-3.5}$. }

{The resulting dust surface density profile is shown in \fg{GMaurprofile}.
The corresponding continuum image for the simulated dust profile is displayed in \fg{gmaur}, in comparison with the observed image of GM Aur \citep{HuangEtal2020}. The orbits of the imposed planet are indicated by the dashed lines and overplotted on top of the observed continuum image. }

{The inward pebble flux is not fully accreted by the planets. The dust surface density grows as the semimajor axis decreases with conserving flux ($\Sigma_\mathrm{d} \sim F/rv_\mathrm{drift}$). Consequently, the simulated GM Aur has an inner disk, which is inconsistent with observation. 
}
\section{Profile of the gap opened by planet}
\label{sec:gapopen}
The gap shape is similar to the one shown in \cite{KanagawaEtal2017}. The radial profile of the surface density in the vicinity of the gap is: 
\begin{equation}
\begin{aligned}
\Sigma_{\mathrm{g}}(r)
     & = \left\{\begin{array}{ll}
        \Sigma_\mathrm{min} & \mathrm{for}\, |r-r_\mathrm{p}|<\Delta r_1, \\
        \Sigma_\mathrm{gap}(r) & \mathrm{for}\,\Delta r_1<|r-r_\mathrm{p}|<\Delta r_2, \\
        \Sigma_0  & \mathrm{for}\, |r-r_\mathrm{p}|>\Delta r_1.
    \end{array}
    \right.
\end{aligned}
\end{equation}
The gas surface density at the bottom of the gap is:
\begin{equation}
    \Sigma_\mathrm{min}=\frac{\Sigma_0}{1+0.04K_\mathrm{3D}},
\end{equation}
where $\Sigma_0$ is the initial gas surface density before the gap opening, and $K_\mathrm{3D}$ is the gap depth factor fitted by 3-D hydrodynamical simulations by \cite{PichierriEtal2024ii}:
\begin{equation}
    K_\mathrm{3D} = 28q^{2.3} h^{-5.4} \alpha^{-0.72}.
\end{equation}
The surface density at the edge of the gap smoothly transitions to $\Sigma_0$ in the form
\begin{equation}
    \frac{\Sigma_\mathrm{gap}(r)}{\Sigma_0}=\frac{4}{\sqrt[4]{K_\mathrm{3D}'}}\frac{|r-r_\mathrm{p}|}{r_\mathrm{p}}-0.32,
\end{equation}
where we scale $K_\mathrm{3D}'=h^2K_\mathrm{3D}$.
The width of the gap excluding and including the transition slop are
\begin{equation}
    \Delta r_1=\left(\frac{\Sigma_\mathrm{min}}{4\Sigma_0}+0.08 \right) r_\mathrm{p},
\end{equation}
and 
\begin{equation}
    \Delta r_2=0.33 \sqrt[4]{K_\mathrm{3D}'} r_\mathrm{p},
\end{equation}
with respectively. 




\section{Quantify the morphology of dust rings}
\label{sec:gaussfit}
For the sample study of transition disks, we selected all known transition disks from \citet{vanderMarel2023} at $<$200 pc with published high-resolution ALMA continuum observations with a spatial resolution of 0.05" or better ($<$10 au resolution), so that the radial profile is sufficiently resolved. Disks with strong asymmetries, high inclinations ($>$70$^{\circ}$ inclination) or cavity size $<$30 au are excluded. This results in 19 targets, listed in \tb{samplefit}, including their beam size and reference to the origin of the data, as well as the parameters of our best fit. Radial intensity profiles are extracted using azimuthal averaging of the images following the original reference, and normalized to the peak.

\begin{figure*}
    \centering
    \includegraphics[width=\textwidth]{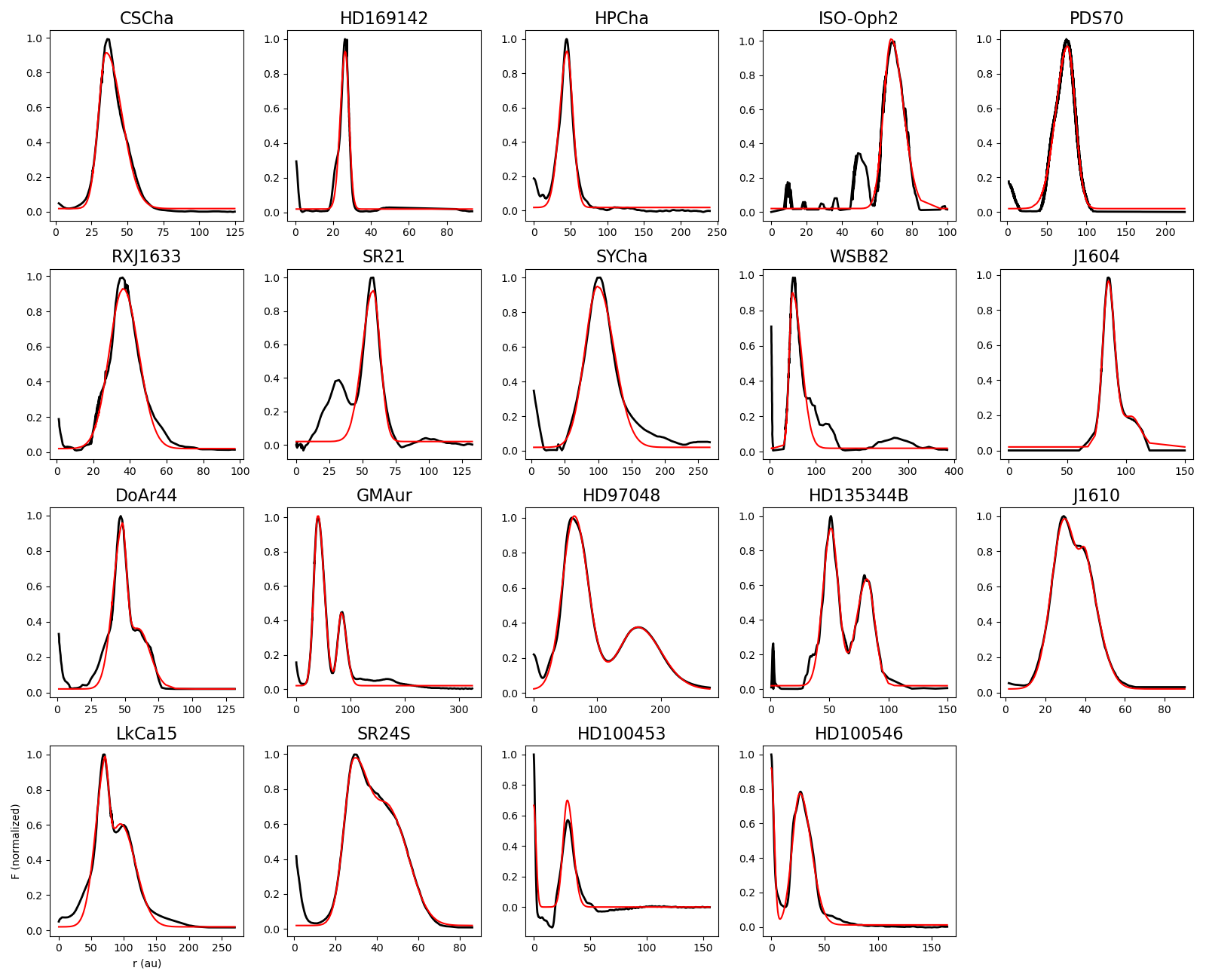}
    \caption{Radial profiles of the selected sample of transition disks. Data are shown in black, best fits in red. The first 9 disks are fit with a single Gaussian, the next 2 disks with a single Gaussian as well as a central peak, and the final 8 disks with two Gaussians, to match both rings. The best-fit parameters of interest are given in \tb{samplefit}.}
    \label{fig:fitting}
\end{figure*}

The radial profiles of each target are fit using a 1D asymmetric Gaussian, or in case of two rings the sum of two 1D asymmetric Gaussians with different peak values. Such a Gaussian is described mathematically as follows:

\begin{equation}
\begin{aligned}
I(r)
     & = \left\{\begin{array}{ll}
        I_0\exp \left(-\frac{(r-r_\mathrm{peak})^2}{2\sigma_\mathrm{left}^2}\right) & {\rm\; for \;} r<r_\mathrm{peak}, \\
        I_0\exp \left(-\frac{(r-r_\mathrm{peak})^2}{2\sigma_\mathrm{right}^2}\right) & {\rm\; for \;} r>r_\mathrm{peak}.
    \end{array}
    \right.
\end{aligned}
\end{equation}

The parameter of interest is the ratio between the outer width and inner width:
\begin{eqnarray}
    R=\frac{\sigma_\mathrm{right}}{\sigma_\mathrm{left}}
\end{eqnarray}
which can distinguish between profiles with a smooth, extended outer profile (like GM Aur) and profiles with a sharp, narrow ring (like PDS 70). When calculating the ratio, we consider an error on the radial width based on the convolution with the beam size, normalized by the number of beams along the dust ring, i.e. $R_\mathrm{err} = R_\mathrm{beam}/\sqrt{N_\mathrm{\rm beams}}$. Although this is a rather simple approach, it is sufficient for our purposes. 

The best fits are determined using the \texttt{curvefit} procedure in \texttt{scipy.optimize}, which is a non-linear least-squares fitting tool. \Fg{fitting} shows the best fits and \tb{samplefit} includes the best-fit parameters as well as the resulting width ratios. The table also lists whether an inner pedestal appears to be present. For HD100453 and HD100546 an inner Gaussian was included in the fit to match the strong centrally peaked emission which is though to originate from free-free emission from the star \citep{Rota2024}.

\begin{table*}
    \centering
    \caption{Gaussian fits of the radial profiles of the sample of transition disks considered in this study.}
    \label{tab:samplefit}
    \begin{tabular}{l|llll|llll|l|ll}
    \hline
        Target & \multicolumn{4}{c|}{Gaussian 1} & \multicolumn{4}{c|}{Gaussian 2} & Ped?$^a$ & \multicolumn{2}{c}{Origin} \\
        &$r_{peak,1}$&$\sigma_{left,1}$&$\sigma_{right,1}$&$R_1$        &$r_{peak,2}$&$\sigma_{left,2}$&$\sigma_{right,2}$&$R_2$ 
        &&beam&Ref.\\
        & (au)&(au)&(au)&&(au)&(au)&(au)&&&(")& \\
        \hline
CSCha	&	35	&	5.4	&	11	&	2.0	$\pm$	0.5	&	-	&	-	&	-	&	-&	N	&	0.05x0.03	&	1	\\
HD169142	&	26	&	2.6	&	2	&	0.8	$\pm$	0.2	&	-	&	-	&	-	&	-	&	Y	&	0.03x0.02	&	2	\\
HPCha	&	45	&	10	&	8	&	0.8	$\pm$	0.1	&	-	&	-	&	-	&	-	&	N	&	0.05x0.03	&	3	\\
ISO-Oph2	&	68	&	4.3	&	6.9	&	1.1	$\pm$	0.1	&	-	&	-	&	-	&	-&	M	&	0.03x0.02	&	4	\\
PDS70	&	76	&	15	&	9.7	&	0.7	$\pm$	0.1	&	-	&	-	&	-	&	-&	Y	&	0.07x0.05	&	5	\\
RXJ1633.9-2442	&	36	&	7.7	&	8.5	&	1.1	$\pm$	0.1	&	-	&	-	&	-	&	-&	M	&	0.02x0.02	&	4	\\
SR21	&	58	&	8.1	&	5.5	&	0.7	$\pm$	0.1	&	-	&	-	&	-	&	-&	M	&	0.04x0.03	&	6	\\
SYCha	&	99	&	19	&	25	&	1.3	$\pm$	0.0	&	-	&	-	&	-	&	-&	N	&	0.04x0.02	&	7	\\
WSB82	&	48	&	6.6	&	20	&	3.0	$\pm$	0.2	&	-	&	-	&	-	&	-&	N	&	0.03x0.02	&	4	\\
HD100453	&	30	&	3.8	&	5.2	&	1.4	$\pm$	0.2	&	-	&	-	&	-	&	-	&	N	&	0.04x0.03	&	3	\\
HD100546	&	27	&	7.1	&	10	&	1.4	$\pm$	0.2	&	-	&	-	&	-	&	-	&	N	&	0.06x0.05	&	3	\\
J16042165	&	85	&	4.9	&	5.4	&	1.1	$\pm$	0.3	&	104	&	10	&	7.7	&	0.8	$\pm$	0.1	&	N	&	0.06x0.04	&	8	\\
DoAr44	&	47	&	4	&	3.8	&	1.0	$\pm$	0.2	&	60	&	10	&	8.9	&	0.9	$\pm$	0.1	&	Y	&	0.03x0.03	&	4	\\
GMAur	&	40	&	7.6	&	11.4	&	1.5	$\pm$	0.2	&	84	&	6.8	&	9.7	&	1.4	$\pm$	0.2	&	N	&	0.05x0.03	&	9	\\
HD97048	&	63	&	19	&	22	&	1.2	$\pm$	0.1	&	167	&	30.1	&	36.9	&	1.2	$\pm$	0.1	&	N	&	0.06x0.03	&	3	\\
HD135344B	&	51	&	6.9	&	6.3	&	0.9	$\pm$	0.2	&	82	&	9.1	&	6.6	&	0.7	$\pm$	0.1	&	Y	&	0.05x0.04	&	10	\\
J16100501	&	30	&	7.1	&	13	&	1.8	$\pm$	0.3	&	41	&	3.3	&	7.3	&	2.2	$\pm$	0.8	&	N	&	0.05x0.04	&	11	\\
LkCa15	&	70	&	15	&	6.4	&	0.4	$\pm$	0.2	&	95	&	19.5	&	23.15	&	1.2	$\pm$	0.1	&	Y	&	0.07x0.05	&	11	\\
SR24S	&	29	&	4.9	&	8	&	1.6	$\pm$	0.2	&	47	&	7.1	&	9.1	&	1.3	$\pm$	0.1	&	N	&	0.04x0.03	&	4	\\
\hline     
    \end{tabular}
$^a$ Presence of a pedestal: Y=Yes, N=No, M=Maybe.\\ 
Refs. 1) \citet{Kurtovic2022}, 2) \citet{Perez2019}, 3) \citet{Francis2020}, 4) \citet{Cieza2021}, 5) \citet{FacchiniEtal2021}, 6) \citet{Yang2023}, 7) \citet{Orihara2023}, 8) \citet{Stadler2023}, 9) \citet{HuangEtal2020}, 10) \citet{Casassus2021}, 11) \citet{FacchiniEtal2020}

\end{table*}

\end{appendix}

\label{lastpage}
\end{document}